\documentclass[prl,aps,twocolumn,groupedaddress,floats,final,superscriptaddress,nopacks]{revtex4-2}
\usepackage{graphicx}
\usepackage[utf8x]{inputenc}
\usepackage{lineno}
\usepackage{color}
\usepackage{subfigure}
\usepackage{latexsym}
\usepackage{epsfig}
\usepackage{psfrag}
\usepackage{amsmath,amsfonts,amssymb,bm,ulem}
\usepackage{float}
\usepackage{ulem}
\usepackage{braket}
\usepackage{hyperref}

\newcommand{\Tr}{\mathrm{Tr}}
\newcommand{\cG}{{\cal G}}

\begin{document}

\title{Strong Coupling Quantum Impurity Solver on the Real and Imaginary Axes}
\author{Yueyi Wang}
\affiliation{Center for Materials Theory, Department of Physics and Astronomy, Rutgers University, Piscataway, NJ 08854, USA}
\author{Kristjan Haule\textsuperscript{\textdagger}}
\affiliation{Center for Materials Theory, Department of Physics and Astronomy, Rutgers University, Piscataway, NJ 08854, USA}

\begin{abstract}
The diagrammatic Monte Carlo method has so far been used mainly for weak-coupling expansions.
Here we show that the strong-coupling expansion offers a key advantage: it can be implemented efficiently on both the real and imaginary frequency axes at finite temperature.
Using a quantum-impurity solver for dynamical mean-field theory (DMFT) as an example, we find rapid convergence with expansion order.
We derive closed-form real-axis Feynman rules for diagrams of arbitrary order, and implement them in a bold hybridization-expansion quantum Monte Carlo (BHQMC) impurity solver.
Benchmarking against state-of-the-art numerical renormalization group (NRG) results for the DMFT Mott transition of the Hubbard model, we obtain a highly accurate frequency-dependent scattering rate at finite temperature.
This enables reliable spectroscopy and provides benchmark transport results within DMFT and cluster-DMFT.
\end{abstract}

\maketitle


\textbf{Introduction.} 
Numerous quantum many-body problems can be solved by expanding the action in powers of a control parameter and summing the resulting perturbative series to high order.
The power of this strategy has been demonstrated by diagrammatic Monte Carlo (DiagMC)~\cite{DiagMC-1,DiagMC-2} in a wide range of systems, including the uniform electron gas~\cite{Chen2019,Haule2022}, the unitary Fermi gas~\cite{DiagMC-2}, the polaron problem~\cite{Polaron}, quantum impurity models~\cite{CTint,CThyb,Haule_prb07c,Ctaux,VDMCDMFT}, and even the Hubbard model in certain regimes~\cite{DiagMC-1,Rossi,PhysRevB.96.041105,PhysRevX.11.011058}. 
Most of these expansions start from a weak-coupling limit, in which the unperturbed action is quadratic and the application of Wick’s theorem is straightforward. By contrast, strong-coupling expansions are less common, because Wick's theorem is absent and the corresponding diagrammatic rules are more involved~\cite{keiter1971diagrammatic,Keiter}. Nevertheless, these rules are well established~\cite{Keiter,Coleman} and have been applied to many problems. A prominent example is the hybridization-expansion continuous-time quantum Monte Carlo (CT-HYB) method~\cite{CThyb,Haule_prb07c}, which has proven to be the most robust numerically exact quantum impurity solver. CT-HYB is formulated in imaginary time, so real-frequency observables require numerical analytic continuation. 
Because analytic continuation is ill-posed and unstable, obtaining accurate real-axis spectra remains challenging. To overcome this issue, real-time formulations have recently been developed for weak-coupling expansions~\cite{LeBlanc2019,Ours,FerreroRT,Ferrero2020}, yielding spectra directly on the real axis without analytic continuation. However, their computational cost grows exponentially with perturbation order: a single imaginary-axis term typically generates at least $2^n$ real-frequency contributions at order $n$.

Using the quantum impurity problem as an example, we show that, in contrast to weak-coupling expansions, the strong-coupling expansion can be carried out nearly as efficiently on the real axis as on the imaginary axis.

We implement a bold strong-coupling expansion on both the imaginary and real axes and demonstrate that it yields highly accurate real-axis spectra, comparable to established state-of-the-art numerical renormalization group (NRG) results. This method is closely related to the conventional hybridization-expansion CT-HYB method~\cite{CThyb,Haule_prb07c}, but uses dressed atomic propagators determined self-consistently.
The computational effort is comparable to that of the bare expansion, but the method is expected to suffer from a much less severe sign problem in cluster DMFT applications because the expansion order is much smaller. 
Crucially, the method provides access to both imaginary- and real-axis observables: the imaginary-axis formulation is well suited for computing thermodynamic quantities such as charge density, total energy, and forces in realistic materials~\cite{Kotliar_rmp06,Force,FreeEnergy}, whereas the real-axis formulation enables precise response-function calculations relevant for spectroscopy.

\smallskip

\noindent \textbf{Action.} 
We start with a generic quantum impurity Hamiltonian consisting of an atomic problem $H_{\rm atm}$ coupled to conduction baths,
\begin{eqnarray}
&&H_{\rm atm}=\sum_{\alpha\beta}\psi^\dagger_\alpha \varepsilon_{\alpha\beta}\psi_\beta + \frac{1}{2}\sum_{\alpha\beta\gamma\delta}U_{\alpha\beta\gamma\delta}\psi^\dagger_\alpha\psi^\dagger_\beta\psi_\gamma\psi_\delta\nonumber\\
&&H_\Delta=\sum_{k,s\alpha}V_{\alpha,k s} \psi^\dagger_\alpha c_{ks} +\textrm{h.c.}+\sum_{k s}\epsilon_{k s} c_{k s}^\dagger c_{k s}.
\end{eqnarray}
Here $\alpha,\beta$ denote combined orbital and spin indices on the impurity, while $k,s$ label the bath states.
The full Hamiltonian is $H=H_{\rm atm}+H_\Delta$.
After integrating out the bath degrees of freedom $c_{ks}$, we obtain the action $S=S_{\rm atm}+S_\Delta$, with the retarded hybridization term
\begin{equation}
S_{\Delta} =\int_0^\beta\int_0^\beta d\tau d\tau'\sum_{\alpha,\beta}\psi^\dagger_\alpha(\tau)\Delta_{\alpha,\beta}(\tau-\tau')\psi_{\beta}(\tau'),
\label{QIM}
\end{equation}
and the local atomic part $S_{\rm atm}$.
Here $\psi$ are Grassmann variables corresponding to the impurity operators introduced above, while the hybridization function is $\Delta_{\alpha\beta}(i\omega_n)=\sum_{ks}
{V_{\alpha,ks}V^*_{\beta,ks}}/
{(i\omega_n-\epsilon_{ks})}$.

In the strong-coupling expansion, we diagonalize the atomic Hamiltonian, $H_{\rm atm}\ket{m}=E_m\ket{m}$, where $E_m$ and $\ket{m}$ are the atomic eigenvalues and eigenstates, respectively.
It is convenient to express $S_{\rm atm}$ in terms of pseudo-particles $a_m$~\cite{Coleman}, which create atomic eigenstates from an auxiliary vacuum, $\ket{m}\equiv a_m^\dagger\ket{0}$~\cite{Coleman,haule2010dynamical}:
\begin{equation}
S_{\mathrm{atm}}=\int_0^\beta d\tau \sum_m a_m^\dagger(\tau)\bigl(\partial_\tau+E_m+\lambda\bigr)a_m(\tau).
\end{equation}
An atomic state with an even (odd) number of electrons is represented by a pseudo-boson (pseudo-fermion)~\cite{Coleman}.
The variable $\lambda$ serves as a pseudo-particle chemical potential and is used to enforce the constraint 
$Q=\sum_m a_m^\dagger a_m=1$, thereby projecting onto the physical Hilbert space.
Consequently, the full action can be rewritten in terms of these pseudo-particles, and the hybridization part takes the form of a retarded interaction:
\begin{eqnarray}
&&  S_{\Delta} = \sum_{\alpha\beta,m n m' n'} \braket{m|\psi_\alpha^\dagger|n}\braket{m'|\psi_\beta|n'}\times
\label{QIM2}\\
 && \int_0^\beta \int_0^\beta d\tau d\tau' a_{m}^\dagger(\tau) a_{n}(\tau)\Delta_{\alpha,\beta}(\tau-\tau') a_{m'}^\dagger(\tau')a_{n'}(\tau').
\nonumber
\end{eqnarray}
In the quantum impurity model, this is a local interaction, while in a more general lattice model it is a nonlocal retarded interaction.
The full action $S=S_{\mathrm{atm}}+S_\Delta$ conserves the pseudo-particle number on each site, but the enlarged Hilbert space also contains unphysical sectors which must be projected out.
An important advantage of the pseudo-particle formulation is that Wick's theorem holds in the grand-canonical space, making the diagrammatic expansion straightforward.
\begin{figure}[t]
  \centering
  \subfigure{\includegraphics[scale=1.0,trim=0cm 4.1cm 6cm 0.25cm, clip]{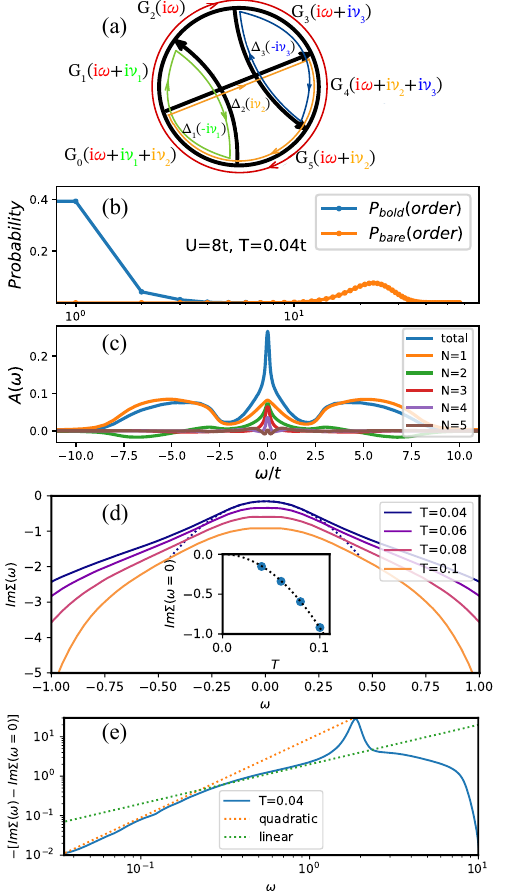}}
\caption{
(a) A third-order Feynman diagram with frequency loops color-coded.
(b) Expansion-order probability distributions for the bare and bold expansions of the action in Eq.~\eqref{QIM2} at $U=8t$ and $T=0.04t$ for the DMFT solution of the 2D square-lattice Hubbard model.
(c) Contributions of different expansion orders to the spectral function for the same parameters.
(d) Imaginary part of the self-energy for the 2D square-lattice Hubbard model within DMFT at $U=9t$ and various temperatures. The dotted line indicates a quadratic fit to the $T=0.04t$ result. Inset: $-\mathrm{Im}\,\Sigma(\omega=0,T)$ and its quadratic fit, yielding $-\mathrm{Im}\,\Sigma(\omega=0,T)/T^2\approx 9.37\pm0.5$.
(e) $-[\mathrm{Im}\,\Sigma(\omega)-\mathrm{Im}\,\Sigma(\omega=0)]$ on a double-logarithmic scale at $T=0.04t$. The dotted lines indicate quadratic and linear frequency dependences.}
\label{Fig2}
\end{figure}

\noindent \textbf{Projection.} 
Because the atomic eigenstates form a complete basis, $\sum_m\ket{m}\bra{m}=1$, the physical Hilbert space contains exactly one pseudo-particle per correlated site at any time. Thus, $Q=\sum_m a_m^\dagger a_m=1$, where $Q$ denotes the pseudo-charge.
The projection onto the $Q=1$ sector of the grand-canonical space is accomplished by introducing the pseudo-particle chemical potential $\lambda$ and taking the limit $\lambda\to\infty$~\cite{Coleman,Abrikosov,supp}.

On the real-frequency axis, if $G_p(\omega)$ denotes a pseudo-particle Green’s function, the spectrum corresponding to charge $Q$ appears at frequencies shifted by $Q\lambda$, i.e., in $G_p(\omega+Q\lambda)$, with $\omega$ of the order of the physical bandwidth.
Since the pseudo-particle and impurity spectra vanish in the $Q=0$ sector (due to the absence of atomic degrees of freedom), taking $\lambda\to\infty$ suppresses all higher-charge sectors, leaving only the physical $Q=1$ contribution.
Since $Q$ corresponds to the number of pseudo-particle loops in a diagram~\cite{Coleman}, only one-loop diagrams with strict time ordering should be considered (example in Fig.~\ref{Fig2}a). 
In practice, the projection in real-axis calculations amounts to shifting all pseudo-particle propagators as $G_p(\omega)\rightarrow G_p(\omega+\lambda)$, while leaving the hybridization function $\Delta(\nu)$ unshifted. 
Further details of the projection are given in Sec. 1 of the Supplemental Material~\cite{supp}.



\noindent \textbf{Simplification of frequency summation.}
As illustrated in Fig.~\ref{Fig2}(a), an order-$n$ one-loop diagram contains $n$ hybridization lines and $n+1$ frequency variables, including $i\omega$ associated with the backbone propagator and $n$ hybridization frequencies $i\nu_m$ carried by $\Delta_m(\pm i\nu_m)$. In a conventional analytic continuation from Matsubara to real frequencies, the number of terms typically grows to at least $2^n$, where $n$ is the diagram order, as each Matsubara sum requires integration along all branch cuts of the propagators in the loop (at least two for each $i\nu_m$). In contrast, the projection in our formulation leads to a drastic simplification: the pseudo-particle self-energy reduces to a single real-axis term, while the impurity Green’s function involves at most $2 n$ terms. As a result, diagram evaluation becomes similarly straightforward on the real and imaginary axes within the Monte Carlo sampling.

This simplification originates from the fact that only one term survives the frequency summation over each hybridization frequency $i\nu_m$. To demonstrate this, we choose the Matsubara frequencies of the hybridization propagators to be fermionic Matsubara frequencies, either $\nu_m$ or $-\nu_m$, where $\nu_m=(2m+1)\pi T$, $m$ is an integer, and $T$ is the temperature (see the color-coded loops in Fig.~\ref{Fig2}(a)). Assume that we have already performed the summations over several $i\nu_m^\alpha$, which have been replaced by integrals over real variables $y_\alpha$, while the summations over $i\nu_l^\beta$ remain to be performed. We now focus on the summation over $i\nu_m$. We have

%
%
{\small\begin{equation}
T\sum_{i\nu_m} \Delta(\pm i\nu_m)
\prod_{j=1}^{N} G_j(i\nu_m+i\omega+\sum_{\alpha} y_\alpha a_{j\alpha}+\sum_\beta i\nu_l^\beta a_{j\beta})
\label{freqsum}
\end{equation}}
where $N$ is the number of pseudo-particle Green’s functions in the $\nu_m$ loop, and $a_{j\alpha},a_{j\beta}\in\{\pm1\}$ encode the frequency flow.


Evaluating the Matsubara sum by contour integration requires enclosing all fermionic Matsubara poles while avoiding the branch cuts of the integrand.
These include the branch cut of $\Delta$ on the real axis and the branch cuts of the pseudo-particle propagators $G_j$, which lie away from the real axis~\cite{supp}.
However, only the branch cut of $\Delta$ survives the projection, yielding the expression:
{\small\begin{equation}
\pm \int dy A_c(\pm y) f(y) \prod_{j=1}^{N} G_j(y+i\omega+\sum_{\alpha} y_\alpha a_{j\alpha}+\sum_\beta i\nu_l^\beta a_{j\beta}),
\label{survived}
\end{equation}}
where $A_c=-\textrm{Im}(\Delta)/\pi$ is the hybridization spectral function and $f$ is the Fermi function. 
We note that this term remains finite and unchanged in the limit $\lambda\rightarrow\infty$. The common frequency $\omega$ in all $G_j$ is shifted by $\lambda$ through the projection and therefore tends to infinity, while the arguments of the hybridization function and the Fermi function remain finite.

In contrast, the contributions from the branch cuts of $G_j$ vanish after projection. Using the spectral representation $G_j(z)=\int dx_j A_j(x_j)/(z-x_j)$, with $A_j$ the pseudo-particle spectral function, a typical term generated by the branch cut of $G_j$ has the form:
\begin{equation}
  \int dx_j A_{j}(x_{j})f(x_{j}-\sum_{\alpha} y_\alpha a_{j\alpha}) R(x_j,i\omega,y_\alpha,i\nu_l^\beta)
\label{notsurvived}
\end{equation}
where $R(\cdots)$ collects the remaining finite factors~\cite{supp}. The crucial observation is that the projection requires shifting $x_j$ by $\lambda$, because it is the argument of the pseudo-particle spectral function $A_j$, while all $y_\alpha$ remain unshifted because they are arguments of the hybridization functions. Consequently, we have $A_j(x_j+\lambda)f(x_j+\lambda-O(1))$, which scales as $e^{-\beta\lambda-\beta x_j}$ and vanishes after projection. Hence, the only surviving term in the sum over $i\nu_m$ is Eq.~\eqref{survived}.

Since the pseudo-particle self-energies contain only $n$ such hybridization-frequency sums, the final real-axis expression always reduces to a single $n$-dimensional integral. The impurity Green’s function requires a summation over $i\omega$, producing at most $2 n$ projected terms, which is negligible compared to the $2^n$ terms encountered in weak-coupling real-time formulations. Further details are provided in the Supplemental Material~\cite{supp}.

\begin{figure}[t]
  \subfigure{\includegraphics[scale=0.7,trim=0.3cm 0.5cm 0.3cm 0.4cm, clip]{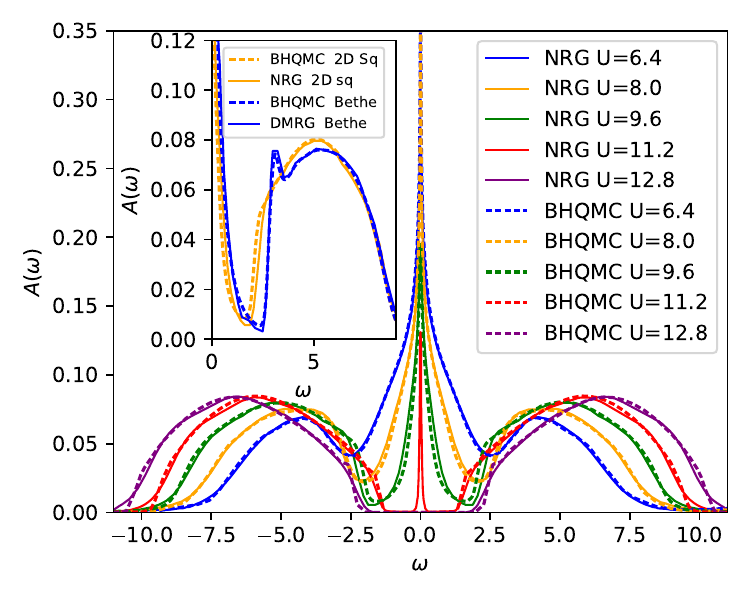}}
  \caption{
  The DMFT spectral function for the 2D square-lattice Hubbard model. Published zero-temperature NRG results from Ref.~\cite{LinearSigma1} (thin lines) are compared with the present BHQMC results at $T=0.04t$ (bold dotted lines). We note that the Mott transition occurs around $U=10t$ at $T=0.04t$~\cite{Park_prl08}, whereas at zero temperature the critical interaction is $U_c=11.6t$~\cite{LinearSigma1}. Inset: Comparison of the spectra for the 2D square and the Bethe lattices at $U=9.6t$, obtained using BHQMC at $T=0.04t$, NRG~\cite{LinearSigma1}, and DMRG~\cite{NRGBethe}.
}
\label{spec}
\end{figure}

\begin{figure}[htbp]
\centering
\subfigure{\includegraphics[scale=0.45, trim=2.1cm 9.2cm 1.3cm 6.5cm, clip]{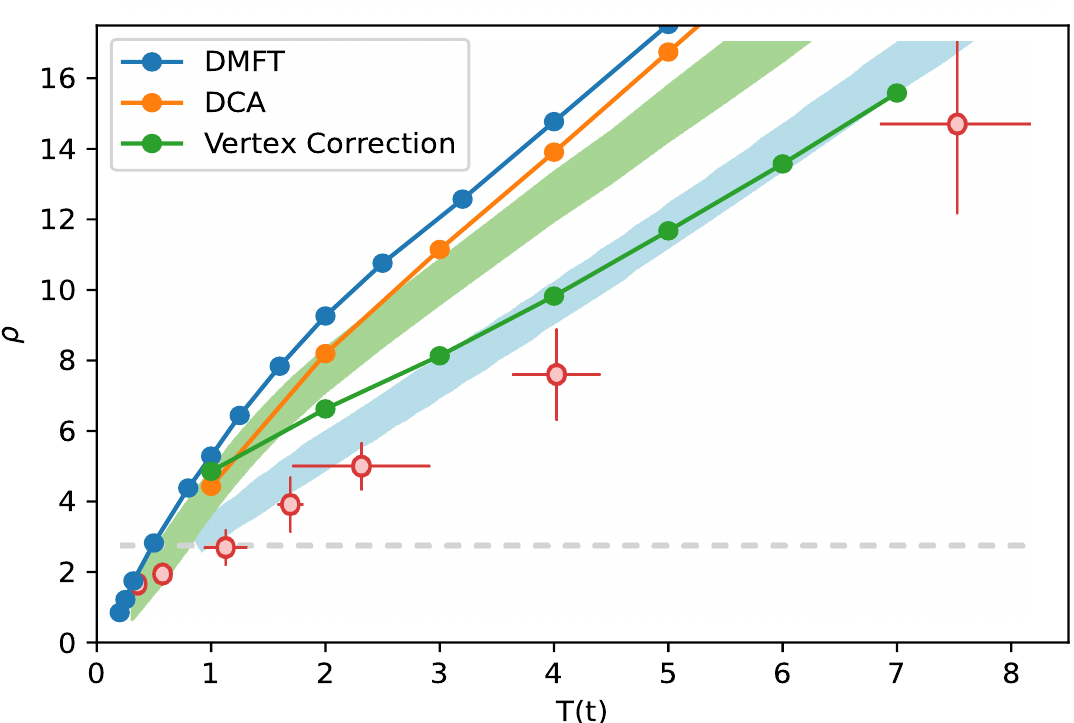}}
\caption{
The resistivity of the 2D Hubbard model at $U=7.5t$ and filling $n_d=0.825$. The cold-atom experimental data (red symbols), finite-temperature Lanczos results (thick blue line), and DMFT results obtained using an exact-diagonalization impurity solver (thick green line) are reproduced from Ref.~\cite{ScienceTR}. 
We compare these results with our BHQMC calculations for single-site DMFT (DMFT), $2\times2$ DCA without current-vertex corrections (DCA), and DCA including the longitudinal vertex correction obtained through the Ward identity (Vertex Correction).
}
\label{resistivity}
\end{figure}

\noindent \textbf{Algorithm.}
We have implemented a diagrammatic Monte Carlo (DiagMC) algorithm that samples, to arbitrarily high order, the Luttinger-Ward functional of the strong-coupling expansion in Eq.~\eqref{QIM2}, retaining only skeleton diagrams.
Our Monte Carlo updates include changes of time/frequency, insertion or removal of a single hybridization line, and exchanges of any two vertices at which hybridization lines begin or end. Most diagrammatic updates generate non-skeleton diagrams, which are skipped until a skeleton diagram is found. This procedure ensures that the space of skeleton diagrams is simply connected and that detailed balance is preserved, as in conventional CT-HYB. The pseudo-particle propagators are fully dressed by the pseudo-particle self-energy obtained by taking the functional derivative of the Luttinger-Ward functional. The physical-electron Green’s function is computed by taking the functional derivative with respect to the hybridization function. 
To improve numerical precision, we compute the physical-electron self-energy from the corresponding two-particle response function using Bulla's trick~\cite{Bulla}, rather than from Dyson's equation (see~\cite{supp}).
Our implementation supports multi-orbital and multi-site impurity models. In the End Matter, we benchmark the method for a two-band Anderson impurity model with a Kanamori interaction and we also apply it to a cluster DMFT calculation in Fig.~\ref{resistivity}.

\noindent \textbf{Results.} Fig.~\ref{Fig2}(b) compares the expansion-order probability distributions of the bare expansion (CT-HYB\cite{CThyb,Haule_prb07c}) and the bold expansion (BHQMC). As illustrated, the bold expansion converges substantially faster than the bare expansion.
The average expansion order $n$ of the bare expansion, which samples the partition function $Z$, is given by $|\Tr(\Delta G)|/T$, i.e., the kinetic energy divided by temperature~\cite{Haule_prb07c}. In contrast, the bold expansion samples the Luttinger-Ward functional, which is proportional to $\log Z$ rather than $Z$ itself. This explains the substantially lower expansion orders observed for the bold expansion in Fig.~\ref{Fig2}(b).
Therefore, for the systems tested here, the bold algorithm is more efficient than the bare expansion, even though the hybridization lines cannot be grouped into determinants.

Fig.~\ref{Fig2}(c) shows the contribution of each expansion order to the electron Green’s function. Higher-order contributions fall off rapidly and become increasingly concentrated near zero frequency. In Fig.~\ref{Fig2}(d), the low-frequency self-energy exhibits Fermi-liquid behavior at sufficiently low temperatures, with the scattering rate showing the expected quadratic dependence on both frequency and temperature. The inset demonstrates the $T^2$ dependence of $-\mathrm{Im}\,\Sigma(0,T)$, while the low-frequency $\omega^2$ behavior is further illustrated on a double-logarithmic scale in Fig.~\ref{Fig2}(e). At intermediate frequencies, $\mathrm{Im}\,\Sigma(\omega)$ crosses over to an approximately linear dependence on frequency.

In Fig.~\ref{spec}, we show the spectral function across the DMFT Mott transition for the 2D square lattice. We observe excellent agreement with NRG over the entire frequency range and for all interaction strengths. The only notable discrepancy occurs at $U=11.2t$: NRG still exhibits a narrow quasiparticle peak, whereas BHQMC predicts an insulating spectrum without a quasiparticle resonance. 
This difference arises because the NRG results are obtained at zero temperature, where $U=11.2t$ remains below the critical interaction $U_c(T=0)=11.6t$~\cite{LinearSigma1}, whereas the BHQMC results are obtained at $T=0.04t$, where $U_c(T=0.04t)\approx10 t$~\cite{Park_prl08}.
Interestingly, the Hubbard bands near the Mott transition ($U=11.2t$) are essentially unchanged between these finite- and zero-temperature calculations. At large $U$, NRG slightly broadens the features near the inner edges of the Hubbard bands, while BHQMC resolves somewhat sharper structures.
For the Bethe lattice, where the side peaks at the inner edges of the Hubbard bands are substantially more pronounced, BHQMC resolves these features with an accuracy comparable to that of the most converged DMRG results obtained by finite-size scaling~\cite{NRGBethe}, as shown in the inset of Fig.~\ref{spec}.

The computation of transport quantities within DMFT is particularly challenging because it requires a highly accurate Green's function on the real axis. Recent advances in cold-atom experiments have enabled measurements of the resistivity of the 2D Hubbard model at high temperatures, allowing direct comparisons with finite-temperature Lanczos calculations and DMFT~\cite{ScienceTR}. Using BHQMC, we recompute the resistivity within single-site DMFT and the $2\times 2$ dynamical cluster approximation (DCA)~\cite{MaierRMP}, and compare our results with earlier estimates based on a finite-size exact-diagonalization impurity solver. The latter are sensitive to the artificial broadening used to smooth the discrete spectral peaks. Our resistivity agrees well with previous results at low temperatures but becomes systematically larger at high temperatures. Our highly accurate results can serve as benchmarks for future developments of quantum impurity solvers.
Although our DMFT resistivity is calculated with high precision, it differs substantially from experiment. This discrepancy arises because current-vertex corrections, which are neglected in single-site DMFT, remain substantial in two dimensions even at elevated temperatures~\cite{Tanaskovic1,Tanaskovic}. To demonstrate that the missing current-vertex corrections are responsible for this discrepancy, we use DCA together with the Ward identity~\cite{Millis_ward,Tanaskovic1,TPSCvertex} to include the longitudinal part of the vertex correction. This correction already captures the correct high-temperature trend, as shown in Fig.~\ref{resistivity}. Further details are provided in the End Matter.

\smallskip

\noindent \textbf{Conclusions.} 
Diagrammatic Monte Carlo has so far been applied primarily to weak-coupling expansions. Here, we have shown that strong-coupling expansions offer two key advantages in the DMFT context: (i) rapid convergence, since the hybridization is a small parameter, and (ii) an efficient real-axis implementation, enabled by closed-form Feynman rules. By implementing a bold hybridization-expansion solver on both the imaginary and real axes, we have demonstrated that finite-temperature spectra can be computed accurately within DMFT without analytic continuation. Benchmark comparisons with NRG highlight the high precision of BHQMC.
Several extensions are now within reach: (a) combining DFT and DMFT for applications to realistic materials; (b) applications to regimes where CT-HYB suffers from a severe fermionic sign problem, such as the $p$-$d$ model of cuprates; and (c) steady-state nonequilibrium problems on the real axis.

\smallskip
\noindent \textbf{Acknowledgements.} We acknowledge support from the NSF under Grants NSF DMR-2233892 and NSF OAC-2311557. Part of this work (the transport calculations) was supported by the U.S. Department of Energy, Office of Science under Award Number \#294274 (K.H).

\bibliography{vDiagMC}

\clearpage
\onecolumngrid
\section{END MATTER}
\twocolumngrid
\begingroup
\setlength{\abovedisplayskip}{4pt plus 1pt minus 1pt}
\setlength{\belowdisplayskip}{4pt plus 1pt minus 1pt}
\setlength{\abovedisplayshortskip}{0pt}
\setlength{\belowdisplayshortskip}{3pt plus 1pt minus 1pt}
\setlength{\jot}{2pt}
\section{Resistivity with vertex corrections}
In the two-dimensional Hubbard model, the resistivity obtained from DMFT is systematically overestimated at high temperatures. This is because current-vertex corrections remain significant in two dimensions but are absent in single-site DMFT. This can be inferred from the fact that cluster extensions of DMFT, when current-vertex corrections are neglected, give a resistivity similar to that of single-site DMFT at high temperatures~\cite{Tanaskovic1}, whereas finite-temperature Lanczos calculations yield a resistivity much closer to the experimental value~\cite{ScienceTR}. Fortunately, part of the vertex correction within cluster extensions of DMFT can be estimated using the Ward identity~\cite{Millis_ward}. Introducing the current vertex $\Gamma_J$, the current-current correlation function can be written as:

\begin{align}
    \chi_{xx}(i\Omega_m)=&\lim_{q \rightarrow 0}
    \sum_{n,\bm k}v_{\bm k_x}G_{\bm{k+q}}(i\omega_n+i\Omega_m)G_{\bm{k}}(i\omega_n) \nonumber \\
    &\times \Gamma_{J}(\bm{k+q},i\omega_n+i\Omega_m;\bm{k},i\omega_n)
    \label{chivertex}
\end{align}
The current vertex $\Gamma_J$ also enters the Ward identity:
\begin{align}
    &\Omega\Gamma_{\rho}(\bm{k}+\bm{q},\omega+\Omega;\bm{k},\omega)-\bm{q}\cdot \Gamma_J(\bm{k}+\bm{q},\omega+\Omega;\bm{k},\omega)\nonumber \\
    &=\Omega-(\epsilon_{\bm{k}+\bm{q}}-\epsilon_{\bm{k}})-[\Sigma(\bm{k}+\bm{q},\omega+\Omega)-\Sigma(\bm{k},\omega)]
\end{align}
Assuming analyticity and inversion symmetry, the expansion of $\Gamma_{\rho}(\bm{k}+\bm{q},\omega+\Omega;\bm{k},\omega)$ 
in powers of $\bm{q}$ 
contains no term linear in $\bm q$.
Taking the long-wavelength and low-frequency limits, the coefficients of $\Omega$ and $\bm q$ can therefore be identified as~\cite{TPSCvertex}:
\begin{align}
    &\Omega\Gamma_{\rho}(\bm{k}+\bm{q},\omega+\Omega;\bm{k},\omega)=\Omega-\Sigma(\bm{k},\omega+\Omega)+\Sigma(\bm{k},\omega) \\
    &\bm q \cdot \Gamma_J(\bm{k}+\bm{q},\omega+\Omega;\bm{k},\omega) \nonumber\\
    &=\epsilon_{\bm{k}+\bm{q}}-\epsilon_{\bm{k}}+\Sigma(\bm{k}+\bm q,\omega+\Omega)-\Sigma(\bm{k},\omega+\Omega).
\end{align}

Here, the momentum-dependent self-energy $\Sigma(\bm k,\omega)$ can be computed using cluster extensions of DMFT, such as DCA~\cite{Millis_ward} or cellular DMFT~\cite{Tanaskovic1}. 
The analytical continuation of $\chi_{xx}$ can be performed by rewriting the right-hand side of Eq.~(\ref{chivertex}) using the spectral representation, summing over $i\omega_n$, and replacing $i\Omega$ with $\Omega+i\delta$:
%
\par
\begin{widetext}
\begin{align}
    \chi^{vertex}_{xx}(\Omega)&=-\frac{1}{\pi}\sum_{\bm k}v_{\bm k_x}\int d\epsilon f(\epsilon)\{G(\bm{k},\epsilon+\Omega)G''(\bm{k},\epsilon) [\Sigma(\bm{k}+\bm q,\epsilon+\Omega)-\Sigma(\bm{k},\epsilon+\Omega)]\\ 
    &+G''(\bm{k},\epsilon)G^*(\bm{k},\epsilon-\Omega)[\Sigma'(\bm{k}+\bm q,\epsilon)-\Sigma'(\bm{k},\epsilon)]
    +G'(\bm{k},\epsilon)G^*(\bm{k},\epsilon-\Omega)[\Sigma''(\bm{k}+\bm q,\epsilon)-\Sigma''(\bm{k},\epsilon)] \},\nonumber 
\end{align}
\end{widetext}
\par
where $G=G'+iG''$, $\Sigma=\Sigma'+i\Sigma''$ and $G^*$ denotes the complex conjugate. A limitation of this approach is that only the longitudinal part of $\Gamma_J$ can be obtained, since the current vertex enters the Ward identity through the dot product $\bm q\cdot\Gamma_J$. The transverse part of the current vertex is generally not negligible, particularly when the self-energy exhibits strong momentum anisotropy or when spin fluctuations dominate the low-energy physics.
However, in the high-temperature, moderately doped regime, we expect the transverse component to provide only a subleading correction to the DC conductivity. Increasing temperature strongly reduces the correlation length and momentum selectivity of collective fluctuations, while moderate hole doping further weakens the incipient antiferromagnetic (nesting-driven) structure associated with half filling.

\begin{figure}[!tb]
\subfigure{\includegraphics[scale=0.45]{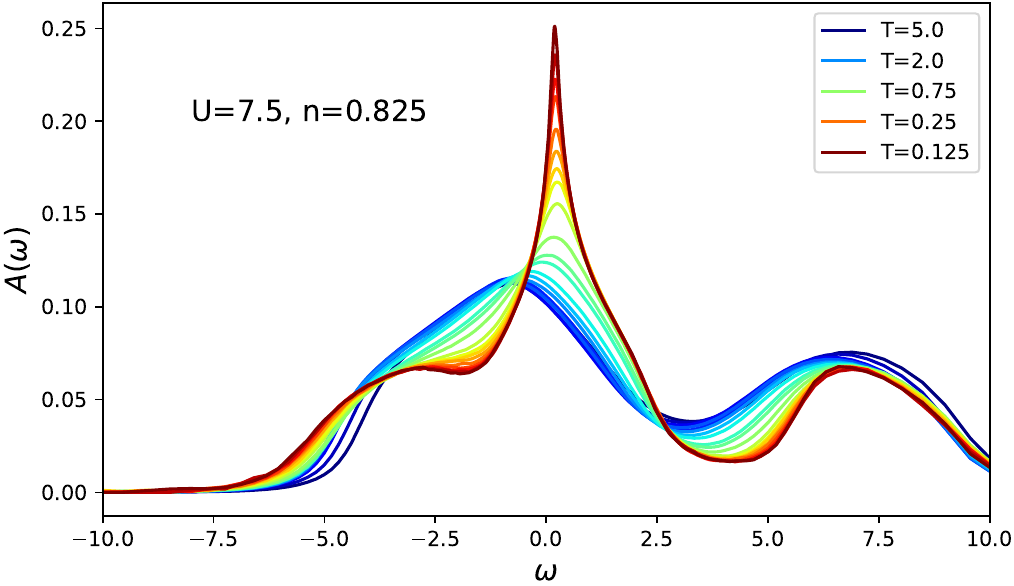}}
\caption{
  The temperature evolution of the single-particle spectral function for $U=7.5t$ and doping $n_d=0.825$.
}
\label{specatdiffT}
\end{figure}

\begin{figure}[!b]
\centering
\includegraphics[scale=0.68]{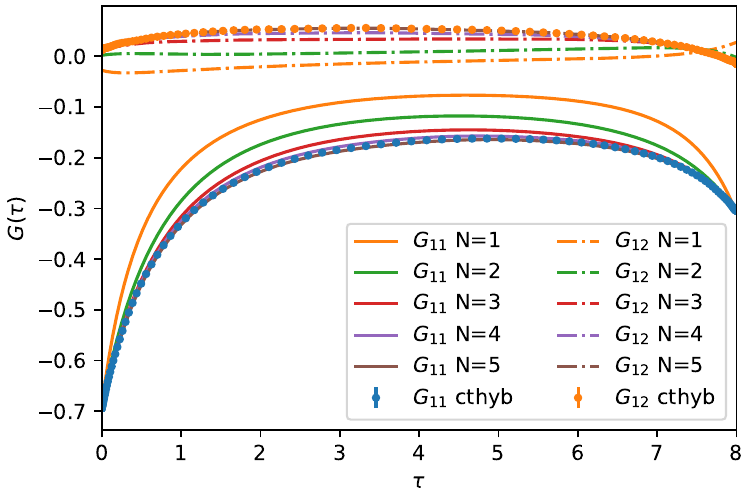}
\caption{The diagonal and off-diagonal Green's functions of the two-band $e_g$ Kanamori model with $r=0.96$, obtained using the imaginary-frequency version of BHQMC and the CT-HYB impurity solver.
}
\label{Gullimag}
\end{figure}
We test this idea in the two-dimensional Hubbard model at hole doping $\delta=0.175$. The single-site DMFT spectra in Fig.~\ref{specatdiffT} show a gradual reduction of the quasiparticle peak with increasing temperature, in contrast to the sudden collapse often produced by analytic continuation of imaginary-axis data. We then compute the DC resistivity within single-site DMFT and $2\times2$ DCA. Without current-vertex corrections, the two give very similar resistivities, showing that short-range correlations in the single-particle self-energy alone do not account for the discrepancy with experiment. We therefore include the longitudinal current vertex $\Gamma_J$ within DCA using the Ward identity. As shown in Fig.~\ref{resistivity}, this correction substantially reduces the resistivity. At very high temperatures, the result agrees with finite-temperature Lanczos calculations and approaches the cold-atom data. At lower temperatures, however, it deviates from the Lanczos results and moves back toward the result without vertex corrections, consistent with the increasing importance of the neglected transverse current vertex.

\section{The two-band $e_g$ model}
To benchmark our multi-orbital implementation against other state-of-the-art impurity solvers, we consider the two-band $e_g$ model studied in Refs.~\cite{JasonPRX,AAAalgorithm,Gullsexample}. The local part of the two-band Anderson impurity model has the Kanamori form
\begin{align}
H_{local}=&U\sum_{i}n_{i\downarrow}n_{i\uparrow}+\sum_{i < j,\sigma\sigma'}(U'-J\delta_{\sigma \sigma'})n_{i\sigma}n_{j\sigma'} \nonumber \\
    &+J\sum_{i \neq j}(c^{\dagger}_{i\uparrow}c^{\dagger}_{j\downarrow}c_{i\downarrow}c_{j\uparrow}+c^{\dagger}_{i\uparrow}c^{\dagger}_{i\downarrow}c_{j\downarrow}c_{j\uparrow})
\label{Kanamori}
\end{align}

where $i,j\in\{1,2\}$ label the orbitals, $U$ is the Hubbard interaction, $U'=U-2J$, and $J$ is the Hund's coupling. We set $U=2t$, $J=0.2t$, $T=t/8$, and the chemical potential $\mu=1$.
The hybridization function is chosen as

\begin{align}
\Delta_{ij}(\omega)=\frac{2}{\pi D^2}\sqrt{D^2-\omega^2}
\begin{pmatrix}
1 & r\\
r & 1
\end{pmatrix}
.
\end{align}

In Refs.~\cite{JasonPRX,AAAalgorithm,Gullsexample} $r=1$ is chosen. 
Here, we focus on the very similar but slightly more generic value $r=0.96$, for which both eigenvalues of the hybridization matrix remain finite after diagonalization.


Fig.~\ref{Gullimag} shows the diagonal and off-diagonal Green's functions obtained with imaginary-frequency BHQMC at different expansion orders, together with the CT-HYB reference results. Consistent with previous work~\cite{AAAalgorithm}, we find that this model requires expansion order $n=5$ to reach quantitative agreement with CT-HYB.

An important advantage of our method is that the same model can also be solved directly on the real-frequency axis, as shown in Fig.~\ref{Gullreal}. To conveniently display the results for the narrow and broad bands, we transform to the bonding-antibonding basis using $\frac{1}{\sqrt 2}\left(\begin{smallmatrix}
1 & 1 \\
1 & -1
\end{smallmatrix}\right)$. 

In this basis, the antibonding orbital couples only weakly to the bath, and its spectral function therefore exhibits two atomic-like peaks with little broadening. In contrast, the bonding orbital couples strongly to the bath and exhibits a continuous spectrum and a smoother self-energy. With our implementation, the fifth-order real-axis results in Fig.~\ref{Gullreal} require approximately 330 core-hours. For comparison, an imaginary-axis algorithm also based on bold sampling requires about 750 core-hours~\cite{AAAalgorithm}, while the inchworm Monte Carlo approach requires about 1500 core-hours~\cite{Gullsexample}.
\par
\onecolumngrid
\par
\begin{figure*}[!b]
\centering
\includegraphics[width=0.97\textwidth]{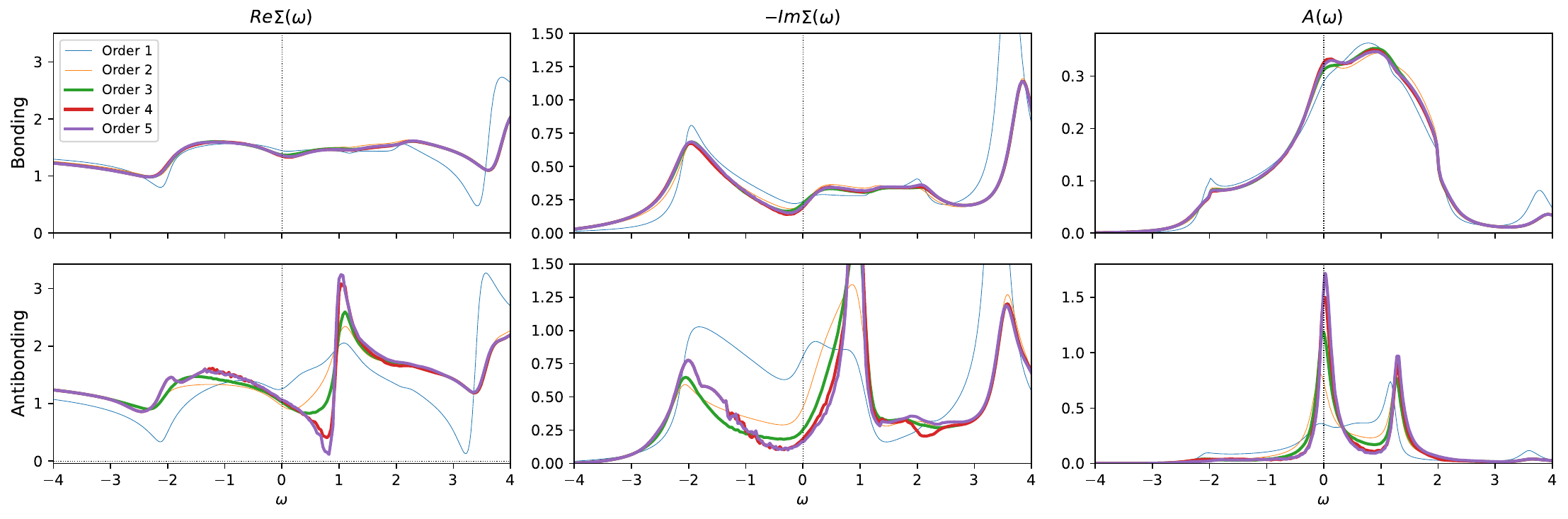}
\caption{The bonding and the antibonding self-energy and spectral functions of the two-band $e_g$ Kanamori model with $r=0.96$, 
obtained using the real-frequency version of the BHQMC impurity solver.
}
\label{Gullreal}
\end{figure*}

\clearpage
\newpage
\newpage
\begin{widetext}
\section{Supplementary information}
\subsection{Projection to the physical Hilbert space}
\label{chap:Projection}


By introducing the pseudo-particle creation operator $a_m^\dagger$ such that each generates an atomic eigen-state $\ket{m}$, i.e., $a_m^\dagger\ket{0}=\ket{m}$, the Hilbert space is substantially expanded. In the physical part of the Hilbert space, only one atomic state is allowed, represented by the constraint $Q=\sum_{m} a_m^\dagger a_m = \sum_m \ket{m}\bra{m}=1$.
To eliminate the unphysical portion of the Hilbert space, an exact projection to the $Q=1$ Hilbert subspace is necessary. This is achieved by introducing a chemical potential for pseudo-particles, i.e., $H\rightarrow H+\lambda Q$. It is then realized that at large $\lambda$, the excitations corresponding to different Hilbert spaces with different $Q$ will be separated by approximately $\lambda$. The physical excitations of $Q=1$ will, therefore, appear at a frequency of $\omega+\lambda$, where $\omega$ is of the order of the bandwidth, and $\lambda$ tends to infinity. Consequently, all frequencies in pseudo-particle propagators need to be shifted by $\lambda$ as $G_p(\omega+\lambda)$.

In addition to shifting the frequency variable  in all pseudo-particle quantities, it is essential to ensure that physical observables are computed by taking the proper trace over the physical Hilbert space. Before the projection, we calculate physical observables in the grand-canonical ensemble, where any value of $Q$ is allowed, and is represented as:
\begin{equation}
  \braket{A} = \frac{\Tr(e^{-\beta (H+\lambda Q)} A)}{\Tr(e^{-\beta (H+\lambda Q)})},
\end{equation}
As $Q$ is a conserved quantity, we can separately perform the trace over states with $Q=0$, $Q=1$, and so on. These traces are denoted as $\Tr_{Q=0}$, $\Tr_{Q=1}$, and so forth. Consequently, we obtain:
\begin{equation}
\braket{A}=  \frac{\Tr_{Q=0}(e^{-\beta H} A)+e^{-\beta\lambda}\Tr_{Q=1}(e^{-\beta H} A)+O(e^{-2\beta\lambda})}{\Tr_{Q=0}(e^{-\beta H})+O(e^{-\beta\lambda})}
\label{aAexp}
\end{equation}
What we are interested in is the trace over the physical Hilbert space, denoted in this notation as 
$$\braket{A}_{Q=1} = \frac{\Tr_{Q=1}(e^{-\beta H} A)}{\Tr_{Q=1}(e^{-\beta H})}.$$
Impurity quantities vanish in the absence of atomic states; therefore, if the observable is the impurity Green's function or the impurity free energy, $\Tr_{Q=0}(e^{-\beta H} A)$ vanishes, hence the first term in the numerator of Eq. (\ref{aAexp}) vanishes. If we consider as the observable charge $Q$, we obtain:
\begin{equation}
  \braket{Q}=\frac{e^{-\beta\lambda}\Tr_{Q=1}(e^{-\beta H})+O(e^{-2\beta\lambda})}{\Tr_{Q=0}(e^{-\beta H})+O(e^{-\beta\lambda})}.
\end{equation}
The average over the physical Hilbert space of an observable, denoted as $\braket{A}_{Q=1}$, can thus be expressed as the ratio of these two quantities:
\begin{equation}
  \braket{A}_{Q=1}=\lim_{\lambda\rightarrow\infty}\frac{\braket{A}}{\braket{Q}}.
\end{equation}
For example, the electron single-particle Green's function $\cG$ must be computed as
\begin{equation}
\cG = \frac{1}{Q} \frac{\delta \Phi}{\delta \Delta}.
\end{equation}
Further details are provided in the next chapter.

\subsubsection{Numerical treatment of projected quantities}

In the next paragraph, we provide some details on the numerical implementation of the projection, which requires special care due to the somewhat different properties of pseudo-particle Green's functions compared to regular electron Green's functions.

We begin with the imaginary axis quantities. The pseudo-particle Green's functions in imaginary time before projection are expressed by standard formulas:
\begin{eqnarray}
G_p(\tau) = \frac{1}{\beta}\sum_{i\omega}  e^{-i\omega\tau} G_p(i\omega) = \int \frac{dx}{\pi} f(-x)e^{-x\tau} G''_p(x+i\delta).
\label{EqGptau}
\end{eqnarray}
This expression is valid for fermionic pseudo-particles, but after projection, the statistics does not matter. Additionally, it is valid only for positive $\tau$. For negative $\tau$, the standard expression is:
$G_p(\tau) =-\int \frac{dx}{\pi} f(x)e^{-x\tau} G''_p(x+i\delta)$, since this is well behaved at $\tau<0$, while the former expression is well behaved for $\tau>0$.
The projection requires shifting the variable $x$ by $\lambda$ to the physical $Q=1$ part of the spectra, as discussed above. For positive times $\tau>0$, we have:
\begin{eqnarray}
  G_p(\tau) = \lim_{\lambda\rightarrow\infty}\int \frac{dx}{\pi} f(-x-\lambda)e^{-x\tau-\lambda\tau} G''_p(x+\lambda+i\delta)=
e^{-\lambda\tau}
  \int \frac{dx}{\pi}
  e^{-x\tau} G''_p(x+\lambda+i\delta)
\label{Eq9}
\end{eqnarray}
which is numerically challenging for large $\lambda$ but is finite at small $\tau$. We note that the same variable shift reveals that for negative times $G_p(\tau<0)\approx - e^{-\lambda\beta}\int \frac{dx}{\pi} e^{-x(\tau+\beta)}G_p''(x+\lambda+i\delta)$
indicating that the Green's function vanishes for any time
$\tau<0$. This means that pseudo-particles do not propagate back in time, and consequently the times on the backbone are time ordered. We also note in passing that choosing bose statistics for pseudo-particle would require one to replace $f(-x)$ in Eq.\ref{EqGptau} by $-n(-x)$, and when the limit $\lambda\rightarrow\infty$ is taken they both give unity, hence statistics of pseudo-particles is irrelevant.

To make numerics stable, we define numerically more appropriate pseudo-particles quantities
\begin{eqnarray}
\widetilde{G}_p(\tau)\equiv e^{\lambda\tau}G_p(\tau).
\end{eqnarray}  
and we will recast all equations into the form that contain only $\widetilde{G}_p(\tau)$, which from Eq.~(\ref{Eq9}), is expressed as:
\begin{eqnarray}
\widetilde{G}_p(\tau)\equiv \int \frac{dx}{\pi}
  e^{-x\tau} G''_p(x+\lambda+i\delta)
\label{Eq11}  
\end{eqnarray}  
We will later prove order by order that $\widetilde{G}_p(\tau)$ is numerically stable quantity.

On the real axis we also encounter a numerical ill-posed problem because at negative values of $x$, the function $G_p''(x+\lambda+i\delta)$ becomes very small, of the order $e^{-|x| \beta}$. To compute physical observables (or, in this case, $\widetilde{G}_p(\tau)$), we must multiply it by a very large number $e^{|x|\tau}$.

To avoid this numerical instability, we devised two different strategies. One commonly used approach in the literature is to employ $G_p''(x+\lambda+i\delta)/f(-x)$. In other words, we rewrite all equations in terms of:
\begin{equation}
\widetilde{G}_p(x) \equiv  G_p''(x+\lambda+i\delta)/f(-x).
\label{Eq12}
\end{equation}
The second strategy, which we implement in Monte Carlo (MC) sampling, involves storing $\log(G_p''(x+\lambda+i\delta))$ in addition to $G_p(x+\lambda+i\delta)$. This is beneficial because, even though $G_p''$ at negative $x$ is of the order $e^{-|x| \beta}$, its logarithm is of the order $-|x|\beta$, which is easy to manage numerically.

We note in passing that $\widetilde{G}_p(\tau)$ and $\widetilde{G}_p(x)$ are interconnected.
Their respective definitions in Eqs.~(\ref{Eq11}) and (\ref{Eq12}) establish the relationship:
\begin{eqnarray}
\widetilde{G}_p(\tau)= \int \frac{dx}{\pi}
  e^{-x\tau} f(-x)\widetilde{G}_p(x).
\end{eqnarray}
This equation represents the analytic continuation kernel for fermionic quantities, indicating that $\widetilde{G}_p(x)$ and $\widetilde{G}_p(\tau)$ share the same spectral density of a fermionic-like quantity in imaginary time or real frequency.
Although the imaginary part of the pseudo-particle spectra behaves as a fermionic quantity, this resemblance is a result of our chosen numerical treatment. 
Pseudo-particles before projection obey either fermionic or bosonic commutation relations (depending on the number of electrons in the atomic state).
After the projection, their statistic does not matter, as both the bosonic and the fermionic pseudo-particles have the same type of response functions.
Additionally, it's worth noting that while the imaginary part of the pseudo-particle Green's functions, when divided by the Fermi function, exhibits femion-like properties, the real part does not. Namely, the real part of the pseudo-particle Green's functions is Kramers-Kronig related to $G_p''(x+\lambda+i\delta)$:
\begin{eqnarray}
G_p'(x) =-\frac{1}{\pi}\int dy \frac{\widetilde{G}_p(y) f(-y)}{x-y}
\end{eqnarray}

Finally we discuss the Dyson equation for pseudo-particles.
It directly follows from the effective action that the Dyson equation on the real axis is given by:
\begin{eqnarray}
  G_m(\omega) = \frac{1}{\omega-E_m-\lambda- \Sigma_m(\omega)}.
\label{DysonR}
\end{eqnarray}
Here, $E_m$ represents the atomic energy levels,  and $\Sigma_m$ is the corresponding pseudo-particle self-energy.
The projection requires shifting the frequency $\omega\rightarrow\omega+\lambda$, resulting in:
\begin{eqnarray}
  G_m(\omega+\lambda) = \frac{1}{\omega-E_m- \Sigma_m(\omega+\lambda)}.
\end{eqnarray}
This form is numerically easy to solve, with the real part computed directly. The imaginary part, on the other hand, can be conveniently computed using:
\begin{eqnarray}
\widetilde{G}_m(\omega)=G''_m(\omega+\lambda)/f(-\omega) = \frac{\widetilde{\Sigma}_m(\omega)}{(\omega-E_m-\Sigma'(\omega+\lambda))^2+(\Sigma''(\omega+\lambda))^2}
\end{eqnarray}
We real axis equations for diagonal pseudo-particle Green's functions, but the generalization to off-diagonal case is completely straighforward.

Solving the Dyson equation on the imaginary axis is more challenging. We first rewrite the Dyson Eq.~\ref{DysonR} as:
\begin{equation}
G = G^0 + G^0 \Sigma G.
\label{DysonT}  
\end{equation}
Here, the product should be understood as the integral over time, and $G^0_{mm}(\tau)=e^{-(E_m+\lambda)\tau}$.
It's important to note that pseudo-particle Green's functions do not allow negative times, as proven above. Hence, we have:
\begin{eqnarray}
G_{mm'}(\tau) = e^{-(E_m+\lambda)\tau}\delta_{mm'} +  \sum_{m''}\int_0^\tau d\tau_2\int_{0}^{\tau_2} d\tau_1 e^{-(E_m+\lambda)(\tau-\tau_2)} \Sigma_{m m''}(\tau_2-\tau_1) G_{m'' m'}(\tau_1)
\end{eqnarray}
This expression is numerically challenging to compute when $\lambda$ is large. However, we introduced above more convenient functions $\widetilde{G}$, in terms of which we can write
\begin{eqnarray}
\widetilde{G}_{mm'}(\tau) = e^{-E_m\tau}\delta_{mm'} +  \sum_{m''}\int_0^\tau d\tau_2\int_{0}^{\tau_2} d\tau_1 e^{-E_m(\tau-\tau_2)} \widetilde{\Sigma}_{mm''}(\tau_2-\tau_1) \widetilde{G}_{m''m'}(\tau_1).
\label{DysonIT}
\end{eqnarray}
Here we used $G_{mm'}(\tau)=e^{-\lambda\tau}\widetilde{G}_{mm'}(\tau)$ and $\Sigma_{mm'}(\tau)=e^{-\lambda\tau}\widetilde{\Sigma}_{mm'}(\tau)$.
Eq.~(\ref{DysonIT})  for $\widetilde{G}_{mm'}$ is called Volterra equation of the second kind, and can be solved iteratively.

We first rewrite Eq.~\ref{DysonIT} into two single integral equations
\begin{eqnarray}
&&\widetilde{G}_{mm'}(\tau)=e^{-E_m\tau}\delta_{mm'} +\sum_{m''}\int_0^\tau C_{m m''}(\tau-t) \widetilde{G}_{m'' m'}(t) dt\label{Eq:X1}\\
&&C_{mm'}(\tau)=\int_0^\tau e^{-E_m(\tau-t)}\widetilde{\Sigma}_{mm'}(t)dt
\label{Eq:X2}
\end{eqnarray}
and we recognize that Eq.~\ref{Eq:X2} can be immediately solved for $C_{mm'}$ once $\widetilde{\Sigma}_{mm'}$ is available. In Eq.~\ref{Eq:X1} $\widetilde{G}_m$ appears on both the left and the right side, hence it looks like an integral equation for $\widetilde{G}$. But this Volterra equation is much simpler to solve than a generic integral equation, because on the right hand side we only need $\widetilde{G}(t<\tau)$ when we are calculating $G(\tau)$. Hence we can proceed iteratively. We know that $\widetilde{G}_{mm'}(\tau=0)=G^0_{mm'}(\tau=0)=\delta_{mm'}$, and with this information we can solve Eq.~\ref{Eq:X1} for the smallest $\tau>0$. And iteratively we can than compute any time $\tau$.

For numerical accuracy, it is convenient to solve for $\Delta G_{mm'}(\tau)=\widetilde{G}_{mm'}(\tau)-e^{-E_m\tau}\delta_{mm'}$. We can cast the above equation into 
\begin{eqnarray}
&& \Delta G_{mm'}(\tau)= R_{mm'}(\tau)+
\sum_{m''}\int_0^\tau dt C_{mm''}(\tau-t)\Delta G_{m'' m'}(t)
\label{Eq:X3}\\
&& R_{mm'}(\tau)=\int_0^\tau C_{mm'}(\tau-t)e^{-E_{m'}t}dt
\label{Eq:X4}
\end{eqnarray}
and start with $\Delta  G(\tau=0)=0$.
Both Eq.~\ref{Eq:X2} and Eq.~\ref{Eq:X4} can be solved by iteration: once we know $C$ at grid point $\tau_i$, we can obtain $C(\tau_{i+1})$ with adding a small part of the integral
\begin{eqnarray}
C_{mm'}(\tau_{i+1}) = C_{mm'}(\tau_i)e^{-(\tau_{i+1}-\tau_i)E_m} + \int_{\tau_i}^{\tau_{i+1}} e^{-(\tau_{i+1}-t) E_m}\Sigma_{mm'}(t)dt\\
R_{mm'}(\tau_{i+1})=R_{mm'}(\tau_i)e^{-(\tau_{i+1}-\tau_i)E_{m'}}+\int_{\tau_{i}}^{\tau_{i+1}} C_{mm'}(t) e^{-E_{m'}(\tau_{i+1}-t)}.
\end{eqnarray}
This update is very cheap and very high accuracy can be achived for integral between two nearby points. We store spline interpolation for $\Sigma_{mm'}(t)$ and $C_{mm'}(t)$ and we can express the above integral of the spline with exponent analytically.

We are hence only left to solve 
Eq.~\ref{Eq:X3}. The complication is that all imaginary axis quanties are best stored on a non-equidistant mesh, because they have exponential drop around $\tau=0$ and exponential rise around $\tau=\beta$. We therefore derive analytic formulas for convolution of two spline functions given on a non-uniform mesh.
For a cubic spline in the interval $\tau_l\le x\le \tau_{l+1}$, the following notation is convenient:
\begin{eqnarray}
C(\tau_l\le x\le \tau_{l+1}) = C^l(x) = c_l \Phi_{l,0}(x)+c_{l+1}\Phi_{l,1}(x)+c_l{''}\Phi_{l,2}(x)+c_{l+1}{''}\Phi_{l,3}(x)
\label{Eq:X99}
\end{eqnarray}
where $c_l$ are values of the spline in the grid-points $\tau_l$ and $c_l{''}$ are second derivatives of the spline. The interpolating functions for the spline are
\begin{eqnarray}
&&\Phi_{l,0}(x)=\frac{\tau_{l+1}-x}{\tau_{l+1}-\tau_l}\\
&&\Phi_{l,1}(x)=\frac{x-\tau_l}{\tau_{l+1}-\tau_l}\\
&&\Phi_{l,2}(x)=\frac{1}{6}(\tau_{l+1}-\tau_l)^2\left[\left(\frac{\tau_{l+1}-x}{\tau_{l+1}-\tau_l}\right)^3-\frac{\tau_{l+1}-x}{\tau_{l+1}-\tau_l}\right]\\
&&\Phi_{l,3}(x)=\frac{1}{6}(\tau_{l+1}-\tau_l)^2\left[\left(\frac{x-\tau_{l}}{\tau_{l+1}-\tau_l}\right)^3-\frac{x-\tau_{l}}{\tau_{l+1}-\tau_l}\right]
\end{eqnarray}
To solve Eq.~\ref{Eq:X3} we write Volterra equation on a non-uniform discreate mesh:
\begin{eqnarray}
\Delta G(\tau_i)-R(\tau_i)=\sum_{j=0}^{j=i-1} \int_{\tau_j}^{\tau_{j+1}}C(\tau_i-t)\Delta G(t)dt=
\sum_{j=0}^{j=i-1} \int_{\tau_j}^{\tau_{j+1}}C(\tau_i-t)\left[\Delta G(\tau_j)\Phi_{j,0}(t)+\Delta G(\tau_{j+1})\Phi_{j,1}(t) \right]
\end{eqnarray}
We here dropped the matrix notation for convenience, $\Delta G_{mm'}\rightarrow \Delta G$,  but restoring the matrix notation is straighforward.
We can rewrite this equation in the following form
\begin{eqnarray}
\Delta G(\tau_i)-R(\tau_i)= \sum_{j=0}^{j=i} \Delta G(\tau_j) W_j^i
\end{eqnarray}
where
\begin{eqnarray}
W_j^i=\int_{\tau_j}^{\tau_{j+1}}C(\tau_i-t)\Phi_{j,0}(t)dt+\int_{\tau_{j-1}}^{\tau_j}C(\tau_i-t)\Phi_{j-1,1}(t)dt
\label{Eq:X5}
\end{eqnarray}
for any $0<j<i$ and the two edge cases at $j=0$ and $j=i$ contain only the first or second term, respectively.
Now, the weights $W^i_j$ can be computed analytically from the knowledge of spline coefficients of function $C(t)$, and then we can solve for $\Delta G(\tau_i)$ with substitution
\begin{eqnarray}
\Delta G(\tau_i)=\frac{1}{1-W_i^i}\left[\sum_{j=0}^{j=i-1} W_j^i \Delta G(\tau_j)+R(\tau_i)\right]
\end{eqnarray}
and the equation can now be solved iteratively.

To compute weights $W^i_j$ from Eq.~\ref{Eq:X5} we first shift the integral variable $x=\tau_i-t$
\begin{eqnarray}
W_j^i=\int^{\tau_i-\tau_j}_{\tau_i-\tau_{j+1}}C(x)\Phi_{j,0}(\tau_i-x)dt+\int^{\tau_i-\tau_{j-1}}_{\tau_i-\tau_j}C(x)\Phi_{j-1,1}(\tau_i-x)dt
\label{Eq:X6}
\end{eqnarray}
and than use definition of a spline interpolation of $C(x)$ from Eq.~\ref{Eq:X99}. We find index $l_0$, $l_1$, and $l_2$ so that the following is satisfied
$\tau_{l_0}<\tau_i-\tau_{j+1}\le \tau_{l_0+1}$ and $\tau_{l_1}<\tau_i-\tau_{j}<\tau_{l_1+1}$, and $\tau_{l_2}<\tau_i-\tau_{j-1}<\tau_{l_2+1}$
and than the weight can be expressed by
\begin{eqnarray}
W^i_j=\int_{\tau_i-\tau_{j+1}}^{\tau_{l_0+1}} C^{l_0}(x)\Phi_{j,0}(\tau_i-x)dx+\sum_{l=l_0+1}^{l=l_1-1}\int_{\tau_l}^{\tau_{l+1}} C^{l}(x)\Phi_{j,0}(\tau_i-x)dx + \int_{\tau_{l_1}}^{\tau_i-\tau_j} C^{l_1}(x)\Phi_{j,0}(\tau_i-x)dx
\nonumber\\
+\int_{\tau_i-\tau_j}^{\tau_{l_1+1}} C^{l_1}(x)\Phi_{j-1,1}(\tau_i-x)+\sum_{l=l_1+1}^{l=l_2-1}\int_{\tau_l}^{\tau_{l+1}}C^l(x)\Phi_{j-1,1}(\tau_i-x)dx+\int_{\tau_{l_2}}^{\tau_i-\tau_{j-1}}C^{l_2}(x)\Phi_{j-1,1}(\tau_i-x)
\end{eqnarray}
Finally, $C^l(x)$ are next expanded as in Eq.~\ref{Eq:X99} and we can analytically integrate polynomials $\int \Phi_{l,t}(x)\Phi_{j,0}(\tau_i-x)dx$ or $\int \Phi_{l,t}(x)\Phi_{j-1,1}(\tau_i-x)dx$ so that integrals are computed exactly within spline interpolation.

\subsection{Method to calculate the electron self-energy}

The electron self-energy is easier to compute from the two particle response function than from the Dyson equation, as first discovered by Bulla in the context of NRG~\cite{Bulla}.
This is particularly crucial here, as the convergence with perturbation order is significantly faster when the self-energy is computed from the two-particle response function. In this case, even the first order (non-crossing approximation method) yields the correct high-frequency behavior. Of course, at high order, both the Dyson equation and the two-particle response function can be used to compute the self-energy, and by comparing the two results, we can also test the convergence.

For generic impurity problem the connection between the electron self-energy $\Sigma(\omega)$ and two particle response function is given by:
\begin{eqnarray}
  ({\cG}(i\omega)\Sigma(i\omega))_{\beta\alpha}=-\sum_{ijk}\int_0^\beta d\tau e^{i\omega\tau}
  \braket{T_\tau \psi_\beta(\tau)\psi^\dagger_i(0)\psi^\dagger_j(0)\psi_k(0^-)}\frac{1}{2}(U_{ijk\alpha}-U_{jik\alpha}).
\label{SGS}
\end{eqnarray}
Here $\cG$ is the electron single-particle Green's function (not to be confused with pseudo-particle Green's function $G_p$), 
and the interaction $\hat{U}$ has the form: $\hat{U}=\sum_{ijk\alpha}U_{ijk\alpha}\psi_i^\dagger \psi_j^\dagger \psi_k\psi_\alpha$. $\Sigma$ here is the electron self-energy.
This expression can be obtained either diagrammatically or using the equation of motion.
Before demonstrating how to compute such a quantity within the current pseudo-particle formulation, we first want to show how the single-particle Green's function $\cG$ is computed, and how it relates to pseudo-particles.

Let's start by revisiting how to compute the electron single-particle Green's function $\cG$ for a generic impurity problem. The impurity partition function is given by:
\begin{eqnarray}
Z = \int {\cal D}[\psi,\psi^+] e^{-S_{\rm atm} - \int_0^\beta\int_0^\beta d\tau d\tau'\sum_{\alpha,\beta}\psi^\dagger_\alpha(\tau)\Delta_{\alpha,\beta}(\tau-\tau')\psi_{\beta}(\tau')},
\label{Eq:impurityZ}
\end{eqnarray}
hence we have
\begin{eqnarray}
  -\frac{\delta \log Z}{\delta \Delta_{\alpha\beta}(\tau-\tau')}=\frac{1}{Z}\int {\cal D}[\psi,\psi^+] e^{-S} \psi^\dagger_\alpha(\tau)\psi_\beta(\tau'),
\end{eqnarray}
which is the expression for the electron Green's function. Therefore we have 
\begin{eqnarray}
{\cG}_{\beta\alpha}(\tau'-\tau) = -\frac{\delta \log Z}{\delta \Delta_{\alpha\beta}(\tau-\tau')}.
\label{GD}
\end{eqnarray}
Eq.~\ref{GD} is used in the conventional CT-HYB (bare expansion). In the bold version, BHQMC, we dress the pseudo-particles with their self-energy, and hence the quantity to sample is not $Z$ but rather the Luttinger-Ward functional $\Phi_\Delta[\{G_p\}]$ in terms of which all self-energies are obtained as $\Sigma_p=\frac{\delta \Phi_\Delta[\{G_p\}]}{\delta G_p}$. This is related to $\log Z$ (before the projection) by the standard Klein functional:
\begin{eqnarray}
-\log(Z) =   \beta\Tr\log(G_p)-\beta\Tr(({G_p^0}^{-1}-{G_p}^{-1}) G_p) + \Phi_\Delta[\{G_p\}]
\label{logZ}
\end{eqnarray}  
Here $G_p$ and $G_p^0$ are the pseudo-particle Green's functions. The Klein functional~\ref{logZ} is stationary with respect to $G_p$, hence $\delta\log(Z)/\delta G_p=0$. The derivative with respect to other quantities, like $\Delta$ is thus
\begin{equation}
  \frac{\delta\log(Z)}{\delta \Delta}=
  \frac{\delta G_p}{\delta \Delta}\frac{\delta\log(Z)}{\delta G_p}+\left(\frac{\partial\log(Z)}{\partial\Delta}\right)_{G_p}=-\left(\frac{\partial\Phi_\Delta}{\partial\Delta}\right)_{G_p}
\label{dZdD}
\end{equation}  
The first term vanishes because $\delta\log(Z)/\delta G_p=0$, and since $G_p^0$ does not depend on $\Delta$, we just need to take the derivative of $\Phi$ functional.

Notice that $\Phi_\Delta[\{G_p\}]$ depends on $\Delta$ in a similar way as $\Phi_U[\{\cG\}]$  depends on interaction $U$ in the in weak coupling theory: the theory is not dressed in this channel. The key difference with the weak coupling is that in the weak coupling the interaction can be dressed with the many-body fluctuations (we can use dressed $W$ instead of $U$ to develop a theory), while here any dressing of $\Delta$ by the Dyson equation vanishes exactly because of the projection. This is because projection requires a single loop of pseudo-particles.

Because the theory is not dressed in $\Delta$, it is obvious from Eqs.~\ref{dZdD} and \ref{GD} that
\begin{equation}
{\cG}_{\beta\alpha}(\tau'-\tau) = \frac{\partial \Phi_{\Delta}(\{G_p\})}{\partial \Delta_{\alpha\beta}(\tau-\tau')},
\end{equation}
where we cut any hybridization lines, but do not cut $G_p$ (treat $G_p$ as a constant).

However, this form is valid in the grand-canonical ensemble (before the projection), and as discussed in chapter \textbf{Projection to the physical Hilbert space}, 
we need to divide this value by $\braket{Q}$ to account for the exact projection. Therefore,
\begin{equation}
{\cG}_{\beta\alpha}(\tau'-\tau) = \frac{1}{\braket{Q}}\frac{\partial \Phi_{\Delta}(\{G_p\})}{\partial \Delta_{\alpha\beta}(\tau-\tau')}.
\end{equation}
In the MC sampling we cut each hybridization line, and we obtain $n$ contribution to the singe-particle Green's function, where $n$ is the perturbation order. Fig.~\ref{SFig1} shows a generic contribution to $\Phi_\Delta$.
\begin{figure}[t]
\subfigure{\includegraphics[scale=0.7]{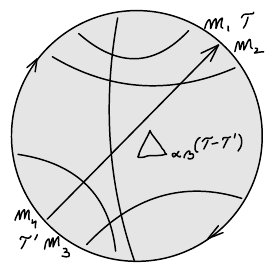}}
\caption{
The sketch of the Luttinger-Ward functional $\Phi_{\Delta}[\{ G_p\}]$ where we concentrate on a single hybridization line $\Delta$.}
\label{SFig1}
\end{figure}
In the pseudo-particle formulation, the single-particle electron Green's function takes the form of the two particle correlation function:
\begin{eqnarray}
\cG_{\beta\alpha}(\tau'-\tau)=-\sum_{\{m\}}\braket{m_4|\psi_\beta|m_3}\braket{m_2|\psi^\dagger_\alpha |m_1}
  \braket{T_\tau a^\dagger_{m_4}(\tau') a_{m_3}(\tau') a^\dagger_{m_2}(\tau) a_{m_1}(\tau) }
\label{Eq:G}
\end{eqnarray}
This  can be easily derived by noting that the action of the annihilation operator on the atomic states is
\begin{eqnarray}
  \psi_\beta=\sum_{\{m\}}\ket{m_4}\braket{m_4|\psi_\beta|m_3}\bra{m_3}\rightarrow
  \sum_{\{m\}}a_{m_4}^\dagger\braket{m_4|\psi_\beta|m_3}a_{m_3}
\end{eqnarray}
and similarly
\begin{eqnarray}
\psi^\dagger_\alpha=\sum_{\{m\}}\ket{m_2}\braket{m_2|\psi^\dagger_\alpha|m_1}\bra{m_1}\rightarrow \sum_{\{m\}}a_{m_2}^\dagger\braket{m_2|\psi^\dagger_\alpha|m_1}a_{m_1}
\end{eqnarray}
We then recognize that each contribution to $\cG$ can be assigned a set of atomic states $m_1,m_2,m_3,m_4$, in which $m_1$ and $m_2$ appear just before and after the action of $\psi^\dagger_\alpha$ operator, and $m_3$ and $m_4$ just before and after $\psi_\beta$ operator, as sketched in Fig.~\ref{SFig1}. We will denote such contribution with
$\left( \frac{\partial\Phi}{\partial\Delta_{\alpha\beta}(\tau-\tau')}\right)_{m_1,m_2\in\tau;m_3,m_4\in\tau'}$ so that
\begin{equation}
  \cG_{\beta\alpha}(\tau'-\tau) = \sum_{\{m\}} \left( \frac{\partial\Phi}{\partial\Delta_{\alpha\beta}(\tau-\tau')}\right)_{m_1,m_2\in\tau;m_3,m_4\in\tau'}.
\end{equation}
This means that
\begin{equation}
 \left( \frac{\partial\Phi}{\partial\Delta_{\alpha\beta}(\tau-\tau')}\right)_{m_1,m_2\in\tau;m_3,m_4\in\tau'}=-\braket{m_4|\psi_\beta|m_3}\braket{m_2|\psi^\dagger_\alpha |m_1}
   \braket{T_\tau a^\dagger_{m_4}(\tau') a_{m_3}(\tau') a^\dagger_{m_2}(\tau) a_{m_1}(\tau) }
\end{equation}

Now we are in a position to rewrite Eq.~\ref{SGS} in pseudo-particle formulation. We first define the imaginary axis equivalent $\cG(i\omega)\Sigma(i\omega)=\int_0^\beta d\tau e^{i\omega\tau} (\cG\cdot\Sigma)(\tau)$  and than the imaginary time quantity is
\begin{eqnarray}
  (\cG\cdot\Sigma)_{\beta\alpha}(\tau'-\tau)=-\sum_{\{m\}}
  \braket{m_4|\psi_\beta|m_3}F^{\alpha}_{m_2,m_1}
  \braket{T_\tau a^\dagger_{m_4}(\tau') a_{m_3}(\tau') a^\dagger_{m_2}(\tau) a_{m_1}(\tau) }
  \nonumber\\
  F^{\alpha}_{m_2,m_1}=\sum_{ijk}\frac{1}{2}(U_{ijk\alpha}-U_{jik\alpha}) \braket{m_2|\psi^\dagger_i\psi^\dagger_j\psi_k|m_1}.
\label{Eq:GS}
\end{eqnarray}
This can be derived by noting that
\begin{eqnarray}
\psi^\dagger_i\psi^\dagger_j\psi_k=\sum_{\{m\}}\ket{m_2}\braket{m_2|\psi^\dagger_i\psi^\dagger_j\psi_k|m_1}\bra{m_1}\rightarrow \sum_{\{m\}}a_{m_2}^\dagger\braket{m_2|\psi^\dagger_i\psi^\dagger_j\psi_k|m_1}a_{m_1}
\end{eqnarray}
We than see that $\cG\cdot\Sigma$ in Eq.~\ref{Eq:GS} requires the same correlation function as $\cG$ in Eq.~\ref{Eq:G}, except that the prefactors are different. We have
\begin{eqnarray}
  (\cG\cdot\Sigma)_{\beta\alpha}(\tau'-\tau)=\sum_{\{m\}} \frac{F^{\alpha}_{m_2,m_1}}{
  \braket{m_2|\psi^\dagger_\alpha |m_1}}
\left(\frac{\partial\Phi}{\partial\Delta_{\alpha\beta}(\tau-\tau')}\right)_{m_1,m_2\in\tau;m_3,m_4\in\tau'}.
\end{eqnarray}
These matrix elements $F^{\alpha}_{m_2,m_1}/\braket{m_2|\psi^\dagger_\alpha |m_1}$ can be precomputed, and hence the contributions to $(\cG\cdot\Sigma)$ is easily computed along with contributions to  $\cG$ with no extra cost.
Once the sampling is concluded, we perform the Fourier transform
of $(\cG\cdot\Sigma)(\tau'-\tau)$
to get $\cG(i\omega)\Sigma(i\omega)$. We compute $\cG(i\omega)$ in the same process, hence $\Sigma(i\omega)$ can be obtained by dividing the two quantities.

On the real axis the equation has a very similar form, namely,
\begin{eqnarray}
  (\cG\cdot\Sigma)_{\beta\alpha}(\omega)&=&\sum_{\{m\}} \frac{F^{\alpha}_{m_2,m_1}}{
  \braket{m_2|\psi^\dagger_\alpha |m_1}}
  \widetilde{\left(\frac{\partial\Phi}{\partial\Delta_{\alpha\beta}(\omega)}\right)}_{m_1,m_2\in\tau;m_3,m_4\in\tau'}\\
   \cG_{\beta\alpha}(\omega)&=&\sum_{\{m\}}
  \widetilde{\left(\frac{\partial\Phi}{\partial\Delta_{\alpha\beta}(\omega)}\right)}_{m_1,m_2\in\tau;m_3,m_4\in\tau'}
\end{eqnarray}
however $\widetilde{\left(\frac{\partial\Phi}{\partial\Delta_{\alpha\beta}(\omega)}\right)}$ is not just the functional derivative, but needs to remove also corresponding Fermi function from the result, as discussed in chapter \textbf{Feynman rules on imaginary axis}.

For completness, we also state the equations for the two particle response functions, i.e., susceptibilities. We start by taking the second derivative of the partition function Eq.~\ref{Eq:impurityZ}:
\begin{eqnarray}
  \chi_{\alpha\beta\gamma\delta}(\tau_1,\tau_2,\tau_3,\tau_4)\equiv  -\braket{T_\tau \psi^\dagger_\alpha(\tau_1)\psi_\beta(\tau_2)\psi^\dagger_\gamma(\tau_3)\psi_\delta(\tau_4)}=
  -\frac{\delta^2 \log Z}{\delta \Delta_{\alpha\beta}(\tau_1-\tau_2)\delta\Delta_{\gamma\delta}(\tau_3-\tau_4)}.
\end{eqnarray}
This is what is used in the bare expansion (conventional CT-HYB). But here we have self-consistent propagators and we want to express them in terms of $\Phi_\Delta$ and $G_p$. We realize that the Dyson equation is always satisfied, therefore, according to Eq.~\ref{logZ} we can take the derivative of $\Phi_\Delta$ functional only
\begin{eqnarray}  \chi_{\alpha\beta\gamma\delta}(\tau_1,\tau_2,\tau_3,\tau_4)&&=\frac{\delta}{\delta\Delta_{\alpha\beta}(\tau_1-\tau_2)}
  \left(\frac{\partial\Phi_\Delta}{\partial\Delta_{\gamma\delta}(\tau_3-\tau_4)}\right)_G
  \label{Eq:Chi}\\
  &&= \left(\frac{\partial^2\Phi_\Delta}{\partial\Delta_{\alpha\beta}(\tau_1-\tau_2)\partial\Delta_{\gamma\delta}(\tau_3-\tau_4)}\right)_G
  +\sum_i
  \left(\frac{\partial^2\Phi_\Delta}{\partial\Delta_{\gamma\delta}(\tau_3-\tau_4) \partial G_i(\tau_{i+1}-\tau_i)}\right)
  \frac{\delta G_i(\tau_{i+1}-\tau_i)}{\delta\Delta_{\alpha\beta}(\tau_1-\tau_2)}
\nonumber
\end{eqnarray}
Here the last equation shows that we might cut two hybridizations, or, just one hybridization and one pseudo-particle propagator. The latter requires the mixed derivative, which plays the central role in this formalism.
The quantity that appears the last in Eq.~\ref{Eq:Chi} can be expressed as follows:
\begin{eqnarray}
  \frac{\delta G_i(\tau_{i+1}-\tau_i)}{\delta\Delta_{\alpha\beta}(\tau_1-\tau_2)}=\int_{\tau_i}^{\tau_{i+1}} d\tau_2'\int_{\tau_i}^{\tau_2'} d\tau_1'
  G_i(\tau_{i+1}-\tau_2')
\frac{\delta\Sigma_i(\tau_2'-\tau_1')}{\delta\Delta_{\alpha\beta}(\tau_1-\tau_2)}  
  G_i(\tau_1'-\tau_i) .
\end{eqnarray}
This follows from taking the derivative of $G G^{-1}=1$ and noticing that $\delta G^{-1}/\delta\Delta=-\delta\Sigma/\delta\Delta$, because $G^{-1}(\omega)=\omega-E-\Sigma$.
Now we realize that the self-energy $\Sigma=\partial\Phi_\Delta/\partial G$, hence its derivative with respect to $\Delta$ is again the mixed derivative, which we introduced before:

\begin{eqnarray}
    \frac{\delta G_i}{\delta \Delta_{\alpha\beta}}=
  \sum_j G_i\left[\left(\frac{\partial \Sigma_i}{\partial \Delta_{\alpha \beta}}\right)_{G}+\frac{\delta G_j}{\delta \Delta_{\alpha \beta}} \left(\frac{\partial \Sigma_i}{\partial G_j}\right)_{\Delta}\right]  G_i  
\end{eqnarray}

where the time indices are omitted for brevity. Here we note $\delta G/\delta \Delta$ appears on both side of the equation above. Solving for $\delta G/\delta \Delta$ gives a form which is similar as the Bethe-Salpeter equation:

\begin{align}
    \frac{\delta G_j}{\delta \Delta}=G_i\left(\frac{\partial \Sigma_i}{\partial \Delta}\right)_G G_i \left( I_{ij}-\left(  \frac{\partial^2\Phi}{\partial G_j\partial G_i}\right)G_iG_i\right)^{-1}
\end{align}

To evaluate all such contributions, we need to sample three quantities:
\begin{eqnarray} \chi^{(1)}(\tau_1,\tau_2,\tau_3,\tau_4)=\left(\frac{\partial^2\Phi_\Delta}{\partial\Delta_{\alpha\beta}(\tau_1-\tau_2)\partial\Delta_{\gamma\delta}(\tau_3-\tau_4)}\right)_G\\ 
\Gamma_{\gamma\delta,p}(\tau_{p+1},\tau_p,\tau_3,\tau_4)=\left(\frac{\partial^2\Phi_\Delta}{\partial\Delta_{\gamma\delta}(\tau_3-\tau_4) \partial G_p(\tau_{p+1}-\tau_p)}\right)\\
\Gamma'_{pr}(\tau_{p+1},\tau_p,\tau_{r+1},\tau_r)=\left(\frac{\partial^2\Phi_\Delta}{\partial G_p(\tau_{p+1}-\tau_p) \partial G_r(\tau_{r+1}-\tau_r)}\right)
\end{eqnarray}  
And with postprocessing, we can compute $\chi=\chi^{(1)}+\chi^{(2)}$, where the second part $\chi^{(2)}$ is:


\begin{eqnarray}
\chi^{(2)}_{\alpha\beta\gamma\delta}= \sum_{pr}
  \Gamma_{\alpha\beta,p}
  G_pG_p(\textbf{I}-G_pG_p\Gamma'_{pr})^{-1}_{pr}\Gamma_{\gamma\delta,r}
\nonumber
\end{eqnarray}  

\subsection{Feynman rules on imaginary axis}
\label{chp:FR}

\begin{figure}[h]
\subfigure{\includegraphics[scale=0.7]{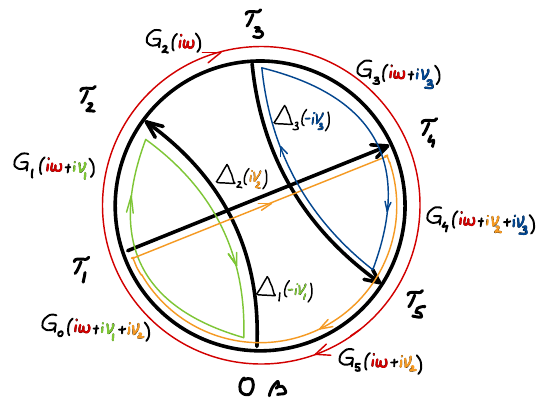}}
\caption{
The sketch of the Luttinger-Ward functional $\Phi_{\Delta}[\{ G_p\}]$.}
\label{SFig2}
\end{figure}
In imaginary time, the Feynman rules are standard.
However, the pseudo-particles $G_p(\tau)$ vanish for negative times. Consequently, the time variables on the backbone increase (appear sorted) and are distributed in the interval $[0,\beta)$. The hybridization functions have a fermionic nature, allowing negative times for hybridization functions, and they exhibit antiperiodic behavior in imaginary time.
To make discussion easier, we show in Fig.~\ref{SFig2} an example
of a typical third-order Feynman diagram for the Luttinger-Ward functional.
Any  self-energy contribution is obtained by cutting the respective propagator.
Without loos of generality, we can choose the first time to be $0$, hence the expression for this particular diagram is
\begin{eqnarray}
  \Phi_{\Delta}[\{G_p\}]=\prod_i\int d\tau_i G_0(\tau_1)  G_1(\tau_2-\tau_1) G_2(\tau_3-\tau_2)G_3(\tau_4-\tau_3)G_4(\tau_5-\tau_4)G_5(\beta-\tau_5)\Delta_1(\tau_2)\Delta_3(\tau_5-\tau_3)\Delta_2(\tau_4-\tau_1),
\nonumber
\end{eqnarray}
where the integral spans the space $0\le\tau_1\le\tau_2\cdots\le\tau_5\le\beta$.

The overall sign could be computed from the number of fermionic loops, following conventional Feynman rules. However, this process is tedious because choosing what constitutes a fermionic loop in these diagrams is not unique. Alternatively, it could be computed by the same method as in conventional CT-HYB. There are two contributions to the sign: i) the parity of permutation of hybridizations, and ii) the matrix elements of the $\psi^\dagger$ and $\psi$ operators.
The parity of permutation for hybridization lines is easy to compute by inspection. In the case of Fig.~\ref{SFig2} the permutation is $(0,2)(1,4)(3,5)$. Here, the numbers in the bracket denotes the start and the end of hybridization line.
Since this is an even permutation of the sorted list $(0,1)(2,3)(4,5)$, the sign contribution (i) is positive.
The matrix elements of $\psi$ operators in Fig.~\ref{SFig2} are $\braket{m_5|\psi_5^\dagger|m_4}\braket{m_4|\psi_4^\dagger|m_3}\braket{m_3|\psi_3|m_2}\braket{m_2|\psi_2^\dagger|m_1}\braket{m_1|\psi_1|m_0}\braket{m_0|\psi_0|m_5}$.
In a single orbital model, only four states of the atom are possible: the empty and doubly occupied states, as well as $\ket{\uparrow}$ and $\ket{\downarrow}$.
If we choose hybridizations $\Delta_1$ and $\Delta_3$ to correspond to $\Delta_\downarrow$ and $\Delta_2$ to $\Delta_\uparrow$ spin, we must have atomic states $\ket{1}$ and $\ket{3}$ to be the empty impurity state, $\ket{5}$ the doubly occupied, and $\ket{0}$ and $\ket{4}$ to be $\ket{\uparrow}$, while $\ket{2}$ is $\ket{\downarrow}$. The product of matrix elements in this case is also $+1$, resulting in an overall sign of $+1$.

To compute the self-energy for the pseudo-particles involved in this functional we need to take the functional derivative in a standard way:
\begin{equation}
  \Sigma_m(\beta-\tau) =\frac{\delta \Phi_\Delta[\{G_p\}]}{\delta G_m(\tau)}
\label{SFD}  
\end{equation}
It's important to note that only positive times are allowed, hence $\Sigma_m(\beta-\tau)$ appears instead of $\Sigma_m(-\tau)$.

Next we discuss how to use numerically stable quantities to evaluate all self-energies. First, all $G_p(\tau)$ should be substituted by $G_p(\tau) =e^{-\lambda\tau}\widetilde{G}_p(\tau)$, where $\widetilde{G}_p$ are numerically stable. It is easy to see that this substitution always gives an overal factor of  $e^{-\beta\lambda}$ to $\Phi_{\Delta}[\{\widetilde{G}\}]$, while otherwise the form of $\Phi_{\Delta}$ remains the same. We will hence define $\widetilde{\Phi}_\Delta \equiv\Phi_\Delta/e^{-\beta\lambda}$ so that $\lambda$ disappears in expression for $\widetilde{\Phi}_\Delta[\{\widetilde{G}\}]$. 
The pseudo-particle self-energies and the pseudo-particle Green's functions also need to be substituted by $\Sigma_p(\tau)=e^{-\lambda\tau}\widetilde{\Sigma}_p(\tau)$. We recognize that in Eq.~(\ref{SFD}) these exponential factors precisely cancel, and hence we can use 
\begin{equation}
  \widetilde{\Sigma}_m(\beta-\tau) =\frac{\delta \widetilde{\Phi}_\Delta[\{\widetilde{G}_p\}]}{\delta \widetilde{G}_m(\tau)}
\label{SFD2}  
\end{equation}
where we removed $\lambda$ from all equations and the projected quantities are now numerically stable. 

Finally, the contribution to the electron single particle Green's function $\cG$, derived above, is
\begin{equation}
\cG_{\beta\alpha}(-\tau) = \frac{1}{\braket{Q}}  \frac{\partial \Phi_\Delta[\{G_p\}]}{\partial \Delta_{\alpha\beta}(\tau)},
\end{equation}
which also requires some adjustments for numeric stability.
As discussed earlier $\Phi_\Delta$ carries a factor of $e^{-\beta\lambda}$
due to the projection to the $Q=1$ subspace.  This factor is canceled by $\braket{Q}$, which is also of the order of $e^{-\beta\lambda}$. More precisely, the occupation of any pseudo-particle is $\braket{a_p^\dagger a_p}=G_p(\beta^-)$, hence
\begin{equation}
\braket{Q} = e^{-\beta\lambda}\sum_p \widetilde{G}_p(\beta).
\end{equation}  
We can thus define $\braket{\widetilde{Q}}=\braket{Q}/ e^{-\beta\lambda}$, and in terms of these quantities, we can write
\begin{equation}
\cG_{\beta\alpha}(-\tau) = \frac{1}{\braket{\widetilde{Q}}}  \frac{\partial \widetilde{\Phi}_\Delta[\{\widetilde{G}_p\}]}{\partial \Delta_{\alpha\beta}(\tau)}
\end{equation}  
where
\begin{eqnarray}
\widetilde{Q}=  \sum_p \widetilde{G}_p(\beta).
\end{eqnarray}  
We thus managed to rewrite all equations in terms of the numerically stable quantities. For the above example we have
\begin{eqnarray}
  \widetilde{\Phi}_{\Delta}[\{\widetilde{G}_p\}]=\prod_i\int d\tau_i \widetilde{G}_0(\tau_1)  \widetilde{G}_1(\tau_2-\tau_1) \widetilde{G}_2(\tau_3-\tau_2)\widetilde{G}_3(\tau_4-\tau_3)\widetilde{G}_4(\tau_5-\tau_4)\widetilde{G}_5(\beta-\tau_5)\Delta_1(\tau_2)\Delta_3(\tau_5-\tau_3)\Delta_2(\tau_4-\tau_1)
\nonumber
\end{eqnarray}
and for example
\begin{eqnarray}
\widetilde{\Sigma}_2(\beta-\tau_3+\tau_2)=\prod_{i\ne 2,3}\int d\tau_{i} \widetilde{G}_0(\tau_1)  \widetilde{G}_1(\tau_2-\tau_1) \widetilde{G}_3(\tau_4-\tau_3)\widetilde{G}_4(\tau_5-\tau_4)\widetilde{G}_5(\beta-\tau_5)\Delta_1(\tau_2)\Delta_3(\tau_5-\tau_3)\Delta_2(\tau_4-\tau_1)
  \nonumber\\
\cG_{3}(\tau_3-\tau_5) =\frac{1}{\braket{\widetilde{Q}}}\prod_{i\ne 3,5}\int d\tau_{i} \widetilde{G}_0(\tau_1)  \widetilde{G}_1(\tau_2-\tau_1) \widetilde{G}_2(\tau_3-\tau_2)\widetilde{G}_3(\tau_4-\tau_3)\widetilde{G}_4(\tau_5-\tau_4)\widetilde{G}_5(\beta-\tau_5)\Delta_1(\tau_2)\Delta_2(\tau_4-\tau_1) 
\nonumber
\end{eqnarray}
Notice that all quantities in these expressions are numerically stable, and Feynman rules are the same as for the original quantities without tilde.

\subsection{Feynman rules on the real axis}
\label{chp:FRR}

Starting from the imaginary-axis expressions for Feynman diagrams, one can always obtain their real-axis counterparts by replacing the Matsubara summations with contour integrals in the complex plane.
However, each Matsubara summation in this step leads to several terms, as many as the number of propagators in the loop.
For a diagram of order $n$, $n$ successive summations are required, and since each loop involves at least two propagators, this generates at least $2^n$ terms on the real axis.
Unfortunately, this often makes Monte Carlo (MC) sampling on the real axis exponentially more expensive than on the imaginary axis.

As explained in the main text, the strong-coupling expansion method offers an important simplification: it generates precisely one term for any Matsubara sum over the hybridization frequency $i\nu$, and exactly $2 n$ terms are generated when evaluating the sum over $i\omega$.
To compute the pseudo-particle self-energy, we need to evaluate $n$ Matsubara sums over the hybridization frequencies $i\nu$, while $i\omega$ becomes the external frequency. As a result, we generate a single term.  On the other hand, the electron Green's function requires $n-1$ Matsubara sums over hybridization frequencies and also the sum over $i\omega$. The former gives a single term, while the latter generates $2 n$ terms. Hence, we end up with $2 n$ terms, which we will derive below. In contrast, in the weak coupling expansion, at least $2^n$ terms are generated.

We will first derive expressions for the pseudo particle self-energies. In this case we need to consider only the integrals over the branch cuts of the hybridization functions, which enormously simplifies the Matsubara summations.
In the main part of the text, we explained that the Matsubara sum over hybridization frequency $i\nu_n$ results in a simple integral over the real variable $y$:
\begin{eqnarray}
&&T\sum_{i\nu_n} \Delta(\pm i\nu_n)\prod_{j=1}^{N} G_j(i\nu_n+i\omega+\sum_{\alpha} y_\alpha a_\alpha^j+\sum_\beta i\nu_m^\beta a_\beta^j)=
\nonumber\\
  &&\pm \int dy A_c(\pm y) f(y) \prod_{j=1}^{N} G_j(y+i\omega+\sum_{\alpha} y_\alpha a_\alpha^j+\sum_\beta i\nu_m^\beta a_\beta^j).
\label{Eq:44}
\end{eqnarray}
This expression follows from the following contour integral in the complex plane:
\begin{eqnarray}
 && T\sum_{i\nu_n} \Delta(\pm i\nu_n)\prod_{j=1}^{N} G_j(i\nu_n+i\omega+\sum_{\alpha} y_\alpha a_\alpha^j+\sum_\beta i\nu_m^\beta a_\beta^j)=
\nonumber\\  
 && -\oint \frac{dz}{2\pi i} \Delta(\pm z) f(z) \prod_{j=1}^{N} G_j(z+i\omega+\sum_{\alpha} y_\alpha a_\alpha^j+\sum_\beta i\nu_m^\beta a_\beta^j).
\label{Eq:45}
\end{eqnarray}
where the contour must enclose all Matsubara points while avoiding all branch cuts of $\Delta$ and $G_j$. 
As explained in the main part of the text, the branch cut on the real axis (at $z=y\in \mathbb{R}$) gives Eq.~\ref{Eq:44}, while we argued that all other branch cuts of $G_j$ give vanishing contributions after projection.
This is because integral around the branch cut of  $G_j$ for $j=p$ requires the substitution of $z\rightarrow y\pm i\delta-i\omega-\sum_{\alpha} y_\alpha a_\alpha^p-\sum_\beta i\nu_m^\beta a_\beta^p$, and we get
$$\int dy A_p(y) f(y-\sum_\alpha a_\alpha^p y_\alpha-\cdots)
\Delta(\pm(y-i\omega-\sum_{\alpha} y_\alpha a_\alpha^p-\sum_\beta i\nu_m^\beta a_\beta^p))\prod_{j=1,j\ne p}^N G_j(y+\sum_{\alpha} y_\alpha (a_\alpha^j-a_\alpha^p)+\sum_\beta i\nu_m^\beta (a_\beta^j-a_\beta^p))
$$
Here $\cdots$ stand for a sum of Matsubara frequencies, which can result in a bosonic or fermionic frequency. If bosonic, we can ignore it in the Fermi function, as $f(x+i\Omega)=f(x)$ ($i\Omega=2 n\pi T$). If fermionic, it gives $f(x+i\Omega)=-n(x)$. But we also apply projection, which requires us to shift the argument of $A_p(y)$ to $A_p(y+\lambda)$ (with $\lambda$ large), so $f(y+\cdots)$ will also shift to $f(y+\lambda+\cdots)=e^{-\beta(y+\lambda+\cdots)}$. Hence, irrespective of the nature of the Matsubara sum denoted by $\cdots$, we get the same exponent $e^{-\beta(y+\lambda+\cdots)}$, with additional minus sign in the prefactor when the Matsubara sum is fermionic. After projection we hence have
\begin{eqnarray}
\pm\int dy A_p(y+\lambda) e^{-\beta(y+\lambda-\sum_\alpha a_\alpha^p y_\alpha)}
  \Delta(\pm(y+\lambda-i\omega-\sum_{\alpha} y_\alpha a_\alpha^p-\sum_\beta i\nu_m^\beta a_\beta^p))\times
\nonumber  \\
\times  \prod_{j=1,j\ne p}^N G_j(y+\lambda+\sum_{\alpha} y_\alpha (a_\alpha^j-a_\alpha^p)+\sum_\beta i\nu_m^\beta (a_\beta^j-a_\beta^p))
\label{Eq:vanish}
\end{eqnarray}
The crucial observation is that all terms in the integral are non-divergent, while the exponent contains a factor $e^{-\beta\lambda}$. Since $\lambda$ is set to infinity, the entire term vanishes after projection.

In Eq.~\ref{Eq:vanish} there are also terms like $e^{\beta a_\alpha^p y_\alpha}$ which become large for large positive $y_\alpha$. However, a more careful derivation shows that for each such term, we also have corresponding Fermi function (from the already performed Matsubara summation), and the combined $f(a_\alpha^p y_\alpha) e^{\beta a_\alpha^p y_\alpha}$ is well-behaved.

We want to point out that if the argument of $G_j$ has the opposite sign, $G_j(-i\nu_n+\cdots)$ then we have to choose $-f(-z)$ instead of $f(z)$ to generate an equally simple result with a single non-vanishing term. In other words:
\begin{eqnarray}
&&T\sum_{i\nu_n} \Delta(\mp i\nu_n)\prod_{j=1}^{N} G_j(-i\nu_n+i\omega+\sum_{\alpha} y_\alpha a_\alpha^j+\sum_\beta i\nu_m^\beta a_\beta^j)=
\nonumber\\
  &&\pm \int dy A_c(\mp y) f(-y) \prod_{j=1}^{N} G_j(-y+i\omega+\sum_{\alpha} y_\alpha a_\alpha^j+\sum_\beta i\nu_m^\beta a_\beta^j).
\label{Eq:57}
\end{eqnarray}
Of course Eq.~\ref{Eq:57} and \ref{Eq:44} are equivalent because $i\nu_n$ go over both the positive and negative frequencies, and the integral over $y$ is unrestricted, allowing a simple substitution, $y\rightarrow -y$.
However, notice that the sign in the Fermi function is tied to the sign in the pseudo Green's function: $f(\pm y) G_j(\pm y+\cdots)$.

Of course, normally, we have the freedom to choose between $f(z)$ and $-f(-z)$, which have the same residues. We just need to be careful to make the resulting expression converge for large $z$. However, here we select the sign in $f(\pm z)$ such that only a single term survives the projection.
For the sake of clarity, lets assume that $G_j$ has a positive argument $G_j(i\nu_n+\cdots)$, like in Eq.~\ref{Eq:44}. Still, we would select $-f(-z)$ instead of $f(z)$, which has the same residues. In this case, several terms would survive the projection, needing to be combined together.  Of course, they would give an identical result, but the derivation in the latter case is more tedious.

In summary, we have just shown that the sign of the Fermi function argument, $f(y)$, in the final results is always equal to the sign in the pseudo-particle Green's function argument, namely, $f(\pm y)G_j(\pm y+...)$ in Eq.~(\ref{Eq:44}).
We now recognize that all Matsubara sums can be carried out by this simple rule. Hence,
the analytic continuation of a Feyman diagram is really simple: we replace $\Delta(i\nu)$ with $A_c(y)f(\pm y)$, where the sign in $f$ argument needs to coinside with the sign of $y$ in pseudo-particle Green's functions.

To make the rules very clear, we provide a concrete form for the Luttinger-Ward functional corresponding to a generic diagram on the imaginary axis:
\begin{eqnarray}
  \Phi \propto T\sum_{i\omega}\left(\prod_{\alpha=1}^n T\sum_{i\nu_\alpha}  \Delta_\alpha(a_\alpha i\nu_\alpha)\right)
  \prod_{j=1}^{2 n} G_j(i\omega+\sum_\beta t_{j,\beta} i\nu_\beta).
\label{Eq:LWe}
\end{eqnarray}  
Here $t_{j,\beta}$ is a set of numbers, which can only take values of either +1,-1 or 0, and $a_\alpha$ is either +1 or -1.
The self-energy contribution to pseudo-particle $G_p$ is a functional derivative, i.e.,
\begin{eqnarray}
\Sigma_p(i\omega) = \frac{\delta\Phi}{\delta G_p(i\omega)}
\end{eqnarray}  
and becomes
\begin{eqnarray}
\Sigma_p(i\omega)\propto \left(\prod_{\alpha=1}^n T\sum_{i\nu_\alpha}  \Delta_\alpha(a_\alpha i\nu_\alpha)\right)
  \prod_{j=1;j\ne p}^{2 n} G_j(i\omega+\sum_\beta t_{j,\beta}^{(p)} i\nu_\beta).
  \label{Eq:60}
\end{eqnarray}  
Here we used a short notation $t_{j,\beta}-t_{p,\beta}\equiv t_{j,\beta}^{(p)}$. As we will demonstrate below $t_{j,\beta}^{(p)}$ also takes the values +1,-1, or 0. Next we slightly manipulate the expression, so that we can use 
Eqs.~\ref{Eq:57} or \ref{Eq:44} on it:
\begin{eqnarray}
\Sigma_p(i\omega)\propto\left(\prod_{\alpha=1}^n T\sum_{i\nu_\alpha}  \Delta_\alpha(a_\alpha i\nu_\alpha)\right)
  \prod_{j=1; j\ne p}^{2 n} G_j(t^{p}_{j,\alpha}i\nu_\alpha+ i\omega+\sum_{\beta\ne\alpha} t^{p}_{j,\beta} i\nu_\beta).
\end{eqnarray}  
Before we carry out the summations, we note that due to projection, only diagrams with a single back-bone survive. As a result, for any $j=1...2n$, the variables $t^{(p)}_{j,\alpha}$ must have the same sign. Specifically, at fixed $\alpha$ and $p$, we either have $t^{(p)}_{j,\alpha}\ge 0$ or $t^{(p)}_{j,\alpha}\le 0$ for any $j$.
This restriction arises because, for a given loop $\alpha$, we can only have either clockwise ($t^{(p)}_{j,\alpha}\ge 0$) or anti-clockwise ($t^{(p)}_{j,\alpha}\le 0$) orientation of the loop through pseudo-particle Green's functions.
To make further derivation easier, we break $t_{j,\alpha}$ into two parts
$t^{(p)}_{j,\alpha}=c^{(p)}_{j,\alpha} b^{(p)}_\alpha$, where $c^{(p)}_{j,\alpha} $ can only take values of 0 or 1, and $b^{(p)}_\alpha$ can only be +1 or -1. Therefore, $b^{(p)}_\alpha$  carries the sign and indicates whether the particular $\alpha$ loop is oriented clockwise or counter-clockwise, while $c^{(p)}_{j,\alpha}=|t^{(p)}_{j,\alpha}|$ is nonzero only for those pseudo-particles $j$ that are in a particular frequency loop $\alpha$.
We can now apply Eqs.~\ref{Eq:57} or \ref{Eq:44} to obtain
\begin{eqnarray}
\Sigma_p(\omega+\lambda)\propto\left(\prod_{\alpha=1}^n a_\alpha b^{(p)}_\alpha\right)\left(\prod_{\alpha=1}^n \int dy_\alpha A^c_\alpha(a_\alpha y_\alpha)f(b^{(p)}_\alpha y_\alpha) \right)
  \prod_{j=1;j\ne p}^{2 n} G_j(\omega+\lambda+i\delta+\sum_{\beta} t^{p}_{j,\beta} y_\beta).
\label{Eq:62}
\end{eqnarray}  
In this equation we also changed $i\omega$ to $\omega+\lambda+i\delta$, which accounts for analytic continuation and projection.
We now notice that appart from overal sign $\prod_{\alpha=1}^n a_\alpha b^{(p)}_\alpha $, which we here determine in alternative way through topology and direct matrix element calculation, the two expressions on real axis (Eq.~\ref{Eq:62}) and imaginary axis (Eq.\ref{Eq:60}) have nearly identical form.
The rule to obtain the real axis expression from the imaginary axis expression is simple:
\begin{itemize}
\item replace $i\nu_\alpha$ with real frequency $y_\alpha$  and replace the Matsubara sum over $i\nu_\alpha$ with integral over $y_\alpha$.
\item replace $i\omega$ with $\omega+\lambda+i\delta$
\item replace hybridization function $\Delta_\alpha$ with its spectral function $A^c_\alpha$.
\item each hybridization function should be accompanied by the Fermi function $f(b^{(p)}_\alpha y_\alpha)$, in which the sign is determined by the sign of the arguments in pseudo-particle Green's functions $b_\alpha=\textrm{sign}(t^{(p)}_{,\alpha})$.
\end{itemize}

To be concrete, we will next discuss the particular diagram in Fig.~\ref{SFig2}. When we calculate the pseudo-particle self-energy $\Sigma_2(\omega)$ the arguments in the pseudo-particle Green's functions are given in table $t^{(2)}_{j,\alpha}$ in Eq.~\ref{tis} and are also denoted in the figure. Each row represents an argument for one of the functions. For example, $G_0$ appears as $G_0(\omega+y_1+y_2)$ once the Matsubara frequencies $i\nu_1$ and $i\nu_2$ are replaced by integration over $y_1$ and $y_2$. However, when we calculate the self-energy $\Sigma_0(\omega)$, we need to shift the arguments by $\omega\rightarrow \omega-y_1-y_2$, which results in arguments displayed in the table $t^{(0)}_{j,\alpha}$. We notice that they can be obtained by the above stated formula $t^{(0)}_{j,\alpha}=t^{(i)}_{j,\alpha}-t^{(i)}_{0,\alpha}$.
The rest of the self-energies require arguments in the tables $t^{(1)}\cdots t^{(5)}$. Now that we know the argument in all $G_j$'s, we can determine the sign of the Fermi functions. We realize that at fixed representation $i$ and frequency loop $\alpha$ all $t^{(i)}_{j,\alpha}$ have the same sign for any $j$, i.e., 
all pseudo-particle Green's functions $G_j$ contain the same sign for argument $y_\alpha$. This sign was before denoted by $b^{(i)}_\alpha=\textrm{sign}(t^{(i)}_{,\alpha})$, and is displayed in Eq.~\ref{Eq:bia}. These numbers $b^{(i)}_\alpha$ give the sign of the arguments in the Fermi functions Eq.~\ref{Eq:62}

\begin{eqnarray}
\left(\begin{array}{c|rrr}
 t^{(2)}& y_1&y_2&y_3\\
\hline
 G_0&1&1&0\\
 G_1&1&0&0\\
 G_2&0&0&0\\
 G_3&0&0&1\\
 G_4&0&1&1\\
 G_5&0&1&0
\end{array}\right)
       \qquad 
\left(\begin{array}{c|rrr}
 t^{(0)}&y_1&y_2&y_3\\
\hline
 G_0&0&0&0\\
 G_1&0&-1&0\\
 G_2&-1&-1&0\\
 G_3&-1&-1&1\\
 G_4&-1&0&1\\
 G_5&-1&0&0
\end{array}\right)
       \qquad 
\left(\begin{array}{c|rrr}
 t^{(1)}& y_1&y_2&y_3\\
\hline
 G_0&0&1&0\\
 G_1&0&0&0\\
 G_2&-1&0&0\\
 G_3&-1&0&1\\
 G_4&-1&1&1\\
 G_5&-1&1&0
\end{array}\right)
 \nonumber\\
\left(\begin{array}{c|rrr}
 t^{(3)}& y_1&y_2&y_3\\
\hline
 G_0&1&1&-1\\
 G_1&1&0&-1\\
 G_2&0&0&-1\\
 G_3&0&0&0\\
 G_4&0&1&0\\
 G_5&0&1&-1
\end{array}\right)
       \qquad 
\left(\begin{array}{c|rrr}
 t^{(4)}& y_1&y_2&y_3\\
\hline
 G_0&1&0&-1\\
 G_1&1&-1&-1\\
 G_2&0&-1&-1\\
 G_3&0&-1&0\\
 G_4&0&0&0\\
 G_5&0&0&-1
\end{array}\right)
       \qquad 
\left(\begin{array}{c|rrr}
 t^{(5)}& y_1&y_2&y_3\\
\hline
 G_0&1&0&0\\
 G_1&1&-1&0\\
 G_2&0&-1&0\\
 G_3&0&-1&1\\
 G_4&0&0&1\\
 G_5&0&0&0
 \end{array}\right)
 \label{tis}         
\end{eqnarray}

$$a=[-1,1,-1]$$
\begin{eqnarray}
\begin{array}{lrrr}  
  b^{(2)} =[&   1,&1,&1]\\
  b^{(0)} =[& -1,&-1,&1]\\
  b^{(1)} =[& -1,&1,&1]\\
  b^{(3)} =[&   1,&1,&-1]\\
  b^{(4)} =[&   1,&-1,&-1]\\
  b^{(5)} =[&   1,&-1,&1]
\end{array}
\label{Eq:bia}                        
\end{eqnarray}

For completness we also show the form of the pseudo-particle self-energies for the example in Fig.~\ref{SFig2}.
We first construct the modified generating functional $\overline{\Phi}$, which contains only hybridizations functions $A^c(y)$ and pseudo-functions $G_j$. The functional form of $\overline{\Phi}$ is identical to its Matsubara imaginary axis form, except that $i\nu_l$ are replaced by real variables $y_l$ and $i\omega$ by $\omega$. For the example in Fig.~\ref{SFig2} the expression is
\begin{eqnarray}
\overline{\Phi} =  \int A^c_1(-y_1) A^c_{2}(y_2) A^c_{3}(-y_3) G_0(\omega+y_1+y_2)G_1(\omega+y_1)G_2(\omega)G_3(\omega+y_3)G_4(\omega+y_2+y_3)G_5(\omega+y_2)
\label{modified}
\end{eqnarray}
Next we add $n$ Fermi functions, one for each hybridization frequency $y_i$, using the above derived rule. The sign of the argument is determined from the tables $b^{(i)}$ and is identical to the sign of that frequency in pseudo-particle Green's function $f(\pm y_i)G_p(\pm y_i+\cdots)$.
We note that the tables $t^{(i)}_{j,\alpha}$ are easily generated ($t^{(i)}_{j,\alpha}=t_{j,\alpha}-t_{i,\alpha}$) for each Feynman diagram once topology is known and frequency loops have been chosen. 
For the above example in Fig.~\ref{SFig2}, we have
\begin{eqnarray}
&&  \Sigma_2(\omega) = \frac{\delta\overline{\Phi}}{\delta G_2(\omega)} f(y_1)f(y_2)f(y_3)\\ &&\Sigma_0(\omega+y_1+y_2)=\frac{\delta\overline{\Phi}}{\delta G_0(\omega+y_1+y_2)} f(-y_1)f(-y_2)f(y_3)\\
&&\Sigma_1(\omega+y_1) = \frac{\delta\overline{\Phi}}{\delta G_1(\omega+y_1)} f(-y_1)f(y_2)f(y_3)\\
&&\Sigma_3(\omega+y_3) = \frac{\delta\overline{\Phi}}{\delta G_3(\omega+y_3)} f(y_1)f(y_2)f(-y_3)\\
&&\Sigma_4(\omega+y_2+y_3) = \frac{\delta\overline{\Phi}}{\delta G_4(\omega+y_2+y_3)} f(y_1)f(-y_2)f(-y_3)\\
&&\Sigma_5(\omega+y_2) = \frac{\delta \overline{\Phi}}{\delta G_5(\omega+y_2)} f(y_1)f(-y_2)f(y_3)
\end{eqnarray}
Here we emphasized that contribution to all pseudo-particle self-energies can be computed from the same form of the functional $\overline{\Phi}$ and we do not actually need to shift variables in $\overline{\Phi}$. This is numerically much more efficient than shifting variables. The shifting of variables was used above only to determine the sign of the Fermi function variables $f(\pm y_\alpha)$.

Finally, we are going to derive the expression for the electron Green's function $\cG(\omega)$. On imaginary axis this is obtained by cutting one of the hybridization propagators, i.e.,
\begin{eqnarray}
\cG(i\nu_\alpha) = \frac{\partial\Phi_{\Delta}[\{G_p\}]}{\partial \Delta_\alpha(i\nu_\alpha)}
\label{Eq:Gdi}
\end{eqnarray}  
As the external variable is now one of the hybridization frequencies, the summation over that particular $i\nu_\alpha$ is omitted,  while the summation over $i\omega$ must be carried out instead.

The end result on the real axis has quite simple structure, but the derivation is somewhat lengthly.
Above we defined $2 n$ matrices $t^{(i)}$, which all represent equivalent ways of choosing frequency arguments in pseudo-particle Green's functions $G_p$, and $t^{(i)}$ stands for representation in which $G_i$ contains only $i\omega$, i.e., $G_i(i\omega)$. Sometimes we omit this superscript $(i)$, when any representation could be chosen, and we did not yet determine which one is being chosen.

Next we want to derive some simple relations between above introduced quantities. We already stated above how to obtain 
representation $t^{(p)}$ from any other representation (say $t^{(l)}$), namely
$t^{(p)}_{j,\alpha} = t^{(l)}_{j,\alpha}-t^{(l)}_{p,\alpha}$. Consequently $t^{(p)}_{p,\alpha}=0$ for any $\alpha$, as required in representation $p$ that $G_p(i\omega)$ has no hybridization frequency $y_\alpha$ in the argument. Second we want to derive $b^{(p)}$ in representation $p$ from knowing $b^{(l)}$ and $c^{(l)}$ in $l$ representation. The connection is simply $b^{(p)}_\alpha = b^{(l)}_\alpha (-1)^{c^{(l)}_{p,\alpha}}$. Note that this comes simply from the algebraic solution of eliminating hybridization frequencies $y_\alpha$ in a chosen pseudo-particle. Also note that because $t^{(p)}_{p,\alpha}={c^{(p)}_{p,\alpha}}=0$ for any $\alpha$ and $p$, the above equation just leads to identity for the case $l=p$.

In Eq.~\ref {Eq:LWe} we stated the Luttinger functional expression on the imaginary axis for a generic Feynman diagram. Using Eq.~\ref{Eq:Gdi}, we obtain
\begin{eqnarray}
  \cG_\alpha(a_\alpha i\nu_\alpha) \propto
T\sum_{i\omega}
\prod_{\beta=1,\beta\ne\alpha}^n T\sum_{i\nu_\beta} 
\Delta_\beta(a_\beta i\nu_\beta)
\prod_{j=1}^{2n} G_j(i\omega+\sum_\beta t_{j,\beta} i\nu_\beta)
\label{imagaG}
\end{eqnarray}  
Here $a_\alpha=\pm 1$ is the relative orientation of the loop
and the hybridization propagator.

Now we start to carry out Matsubara summations. First we perform $n-1$ summations over the hybridization frequencies $i\nu_\beta$, which were explained above, and give a single term on the real axis, which takes the form
\begin{eqnarray}
  \cG_\alpha(a_\alpha i\nu_\alpha) \propto \prod_{\beta=1,\beta\ne\alpha}^n a_\beta b_\beta\int dy_\beta 
  A^c_\beta(a_\beta y_\beta) f(b_\beta y_\beta)
\;T\sum_{i\omega}  \prod_{j=1}^{2n} G_j(i\omega+\sum_{\beta\ne\alpha} t_{j,\beta} y_\beta+t_{j,\alpha} i\nu_\alpha).
\end{eqnarray}  

Next we sum over $i\omega$, which can be either fermionic or bosonic. We first rewrite all pseudo-particle Green's functions in their spectral representation
$G_j(z)=\int dx_j A(x_j)/(z-x_j)$, which gives
\begin{eqnarray}
  \cG_\alpha(a_\alpha i\nu_\alpha) \propto
  \prod_{\beta=1,\beta\ne\alpha}^n a_\beta b_\beta\int dy_\beta 
  A^c_\beta(a_\beta y_\beta) f(b_\beta y_\beta)
\;T\sum_{i\omega}  \prod_{j=1}^{2n}\int dx_j A_j(x_j) \frac{1}{i\omega+\sum_{\beta\ne\alpha} t_{j,\beta} y_\beta+t_{j,\alpha} i\nu_\alpha-x_j}
\end{eqnarray}  
Now we recognize that for this summation we can use the generalized residue formula, which can for example be found in Ref.~\cite{Konik},
\begin{eqnarray}
T\sum_{i\omega}\prod_{j=1}^{N}\frac{1}{i\omega+z_j}=\sum_{l=1}^N f(-z_l)\prod_{j=1,j\ne l}^{N}\frac{1}{z_j-z_l}
\end{eqnarray}
Here $f(z)$ is the Fermi function $f(z)=1/(\exp(\beta z)+1)$ or minus bose function $f(z)=-1/(\exp(\beta z)-1)$ if 
$i\omega$ is fermionic or bosonic, respectively. 
In our case $z_j=\sum_{\beta\ne\alpha} t_{j,\beta} y_\beta+t_{j,\alpha} i\nu_\alpha-x_j$, hence the result is
\begin{eqnarray}
&&  \cG_\alpha(a_\alpha i\nu_\alpha) \propto
  \prod_{\beta=1,\beta\ne\alpha}^n a_\beta b_\beta\int dy_\beta
  A^c_\beta(a_\beta y_\beta) f(b_\beta y_\beta)
\nonumber\\  
&&  \sum_{l=1}^{2n} \int dx_l A_l(x_l) f(x_l-\sum_{\beta\ne\alpha}t_{l,\beta} y_\beta-t_{l,\alpha} i\nu_\alpha)
\prod_{j=1,j\ne l}^{2n} \int dx_j
  \frac{ A_j(x_j)}{\sum_{\beta\ne\alpha} (t_{j,\beta}-t_{l,\beta}) y_\beta+(t_{j,\alpha}-t_{l,\alpha}) i\nu_\alpha+x_l-x_j}.
\end{eqnarray}  
Now we rewrite the spectral representation of $G_j$ back into their closed form for $j\ne l$ terms, to obtain
\begin{eqnarray}
&&  \cG_\alpha(a_\alpha i\nu_\alpha) \propto
  \prod_{\beta=1,\beta\ne\alpha}^n a_\beta b_\beta\int dy_\beta
  A^c_\beta(a_\beta y_\beta) f(b_\beta y_\beta)
\nonumber\\  
&&  \sum_{l=1}^{2n} \int dx_l A_l(x_l) f(x_l-\sum_{\beta\ne\alpha}t_{l,\beta} y_\beta-t_{l,\alpha} i\nu_\alpha)
   \prod_{j=1,j\ne l}^{2n}
G_j(x_l+t^{(l)}_{j,\alpha}i\nu_\alpha+\sum_{\beta\ne\alpha}t^{(l)}_{j,\beta} y_\beta),
\end{eqnarray}
where we used $t^{(l)}_{j,\alpha}=t_{j,\alpha}-t_{l,\alpha}$.
We introduced before the split of $t_{l,\alpha}$ into $b$ and $c$, i.e., $t_{l,\alpha}=b_\alpha c_{l,\alpha}$, where $c_{l,\alpha}=|t_{l,\alpha}|$, hence $e^{-\beta i\nu_\alpha t_{l,\alpha}}=(-1)^{c_{l,\alpha}}$, because external Matsubara frequency $i\nu_\alpha$ is fermionic. Because of projection, we also need to shift $x_l$ frequency to $x_l+\lambda$, hence the $f$ function becomes
$$f(x_l-\sum_{\beta\ne\alpha}t_{l,\beta} y_\beta-t_{l,\alpha} i\nu_\alpha)\rightarrow \pm (-1)^{c_{l,\alpha}}\exp\left(-\beta(x_l+\lambda-\sum_{\beta\ne\alpha}t_{l,\beta} y_\beta)\right).$$ Here $+1$ and $-1$ corresponds to $i\omega$ being fermionic or bosonic. We will therefore introduce notation $(-1)^{\omega_{boson}}$.
After the frequency $x_l$ is shifted by $\lambda$ to account for projection, we obtain
\begin{eqnarray}
&&  \cG_\alpha(a_\alpha i\nu_\alpha) \propto
  (-1)^{\omega_{boson}}\prod_{\beta=1,\beta\ne\alpha}^n a_\beta b_\beta\int dy_\beta
  A^c_\beta(a_\beta y_\beta) f(b_\beta y_\beta) 
\nonumber\\  
  &&  \sum_{l=1}^{2n} e^{\beta\sum_{\beta\ne\alpha} t_{l,\beta} y_\beta}(-1)^{c_{l,\alpha}}\int dx_l A_l(x_l+\lambda)
 e^{-\beta(x_l+\lambda)}   
   \prod_{j=1,j\ne l}^{2n}
G_j(x_l+\lambda +t^{(l)}_{j,\alpha} i\nu_\alpha+\sum_{\beta\ne\alpha}t^{(l)}_{j,\beta} y_\beta)
\end{eqnarray}  
We next combine the term $e^{\beta\sum_{\beta\ne\alpha} t_{l,\beta} y_\beta}$ with the Fermi functions under the product $\prod_{\beta\ne\alpha} f(b_\beta y_\beta)$. We defined above that $t_{l,\beta}=c_{l,\beta}b_\beta$ and $c_{l,\beta}$ is either 0 or 1. We use the fact that $f(x)e^{\beta x}=f(-x)$ to write
$$f(b_\beta y_\beta)e^{\beta y_\beta t_{l,\beta} }=f(b_\beta y_\beta (-1)^{c_{l,\beta}}).$$ 
This now gives
\begin{eqnarray}
&&  \cG_\alpha(a_\alpha i\nu_\alpha) \propto   \sum_{l=1}^{2n}(-1)^{c_{l,\alpha}+\omega_{boson}}\prod_{\beta=1,\beta\ne\alpha}^n a_\beta b_\beta\int dy_\beta
   A^c_\beta(a_\beta y_\beta) f(b_\beta y_\beta(-1)^{c_{l,\beta}})
 \nonumber\\
  &&
     \int dx A_l(x+\lambda)
 e^{-\beta(x+\lambda)}   
   \prod_{j=1,j\ne l}^{2n}
G_j(x+\lambda +t^{(l)}_{j,\alpha} i\nu_\alpha+\sum_{\beta\ne\alpha}t^{(l)}_{j,\beta} y_\beta)
\end{eqnarray}  
%
Next we use the fact $b_\alpha (-1)^{c_{l,\alpha}}=b^{(l)}_\alpha$, and $b_\alpha^2=1$ and $a_\alpha^2=1$, therefore
the sum over $l$ in the above equation is the sum over $2 n$ representations $t^{(l)}$. We have
\begin{eqnarray}
&&  \cG_\alpha(a_\alpha i\nu_\alpha) \propto \left((-1)^{\omega_{boson}}\prod_{\beta=1}^n a_\beta b_\beta\right)
  \sum_{l=1}^{2n}a_\alpha b^{(l)}_\alpha \prod_{\beta=1,\beta\ne\alpha}^n \int dy_\beta
   A^c_\beta(a_\beta y_\beta) f(b^{(l)}_\beta y_\beta)
 \nonumber\\
  &&
     \int dx A_l(x+\lambda)
 e^{-\beta(x+\lambda)}   
   \prod_{j=1,j\ne l}^{2n}
G_j(x+\lambda +t^{(l)}_{j,\alpha} i\nu_\alpha+\sum_{\beta\ne\alpha}t^{(l)}_{j,\beta} y_\beta)
\end{eqnarray}  
Next we perform analytic continuation and let $i\nu_\alpha\rightarrow y_\alpha+i\delta$. We will generate $G_j(\cdots)$, $\textrm{Re}(G_j(\cdots))$, or $G^*_j(\cdots)$ if the coefficient in front of $i\nu$ is 1, 0, or -1. We will therefore define
\begin{eqnarray}
&&  G_j^{(0)}(x) \equiv \textrm{Re}(G_j(x))\\
&&  G_j^{(1)}(x)\equiv G_j(x)\\
&&  G_j^{(-1)}(x)\equiv G_j^*(x)\\
&& \widetilde{A}_j(x) = A_j(x) e^{-\beta x}  
\end{eqnarray}
and in terms of these pseudo-particle Green's function, we have
\begin{eqnarray}
  &&  \cG_\alpha(a_\alpha y_\alpha) \propto
     \left((-1)^{\omega_{boson}}\prod_{\beta=1}^n a_\beta b_\beta\right)
  \sum_{l=1}^{2n}a_\alpha b^{(l)}_\alpha\prod_{\beta=1,\beta\ne\alpha}^n \int dy_\beta
   A^c_\beta(a_\beta y_\beta) f(b^{(l)}_\beta y_\beta)
 \nonumber\\
  &&
     \int dx \widetilde{A}_l(x+\lambda)
   \prod_{j=1,j\ne l}^{2n}
G^{(a_\alpha t^{(l)}_{j,\alpha})}_j(x+\lambda +\sum_{\beta}t^{(l)}_{j,\beta} y_\beta)
\end{eqnarray}
The first bracket gives just an overal sign, which needs to be combined with other signs (like number of fermionic loops, and perturbation order) that we ignored at the beginning, because we determine them in alternative way, i.e., in the same way as in conventional CT-HYB.
We will use notation
\begin{eqnarray}
  (-1)^{sg}\equiv (-1)^{\omega_{boson}}\prod_{\beta=1}^n a_\beta b_\beta.
\end{eqnarray}
It can be shown that this sign is independent of the choice of the loops (the choice of representation) and is a property of the topology of diagram.

Finally, we got the following expression for the electron Green's function
\begin{eqnarray}
&&  \cG_\alpha(a_\alpha y_\alpha) \propto
   (-1)^{sg}\sum_{l=1}^{2n} a_\alpha b^{(l)}_\alpha
   \prod_{\beta=1,\beta\ne\alpha}^n \int dy_\beta
   A^c_\beta(a_\beta y_\beta) f(b^{(l)}_\beta y_\beta)
 \nonumber\\
  &&
     \int dx \widetilde{A}_l(x+\lambda)
   \prod_{j=1,j\ne l}^{2n}
G^{(a_\alpha t^{(l)}_{j,\alpha})}_j(x+\lambda +\sum_{\beta}t^{(l)}_{j,\beta} y_\beta)
\end{eqnarray}
While this is a closed expression that could be coded, it is numerically more convenient to compute all Green's function in the $l$ sum using the same set of arguments. To get that, we need to transform from $(l)$ representation back to the original representation in which the diagram was initially constructed. We will therefore shift the frequency $x$ to $x= \omega+\sum_\beta t_{l,\beta}$ to obtain
\begin{eqnarray}
\cG_\alpha(a_\alpha y_\alpha) \propto &&
   (-1)^{sg}\sum_{l=1}^{2n}a_\alpha b^{(l)}_\alpha
   \int d\omega \prod_{\beta=1,\beta\ne\alpha}^n \int dy_\beta
   A^c_\beta(a_\beta y_\beta) f(b^{(l)}_\beta y_\beta)
 \nonumber\\
  &&
      \widetilde{A}_l(\omega+\lambda+\sum_\beta t_{l,\beta})
   \prod_{j=1,j\ne l}^{2n}
     G^{(a_\alpha t^{(l)}_{j,\alpha})}_j(\omega+\lambda +\sum_{\beta}t_{j,\beta} y_\beta)
\label{Eq:reG}     
\end{eqnarray}
so that all pseudo-particle Green's functions can be evaluated in any representation in which the diagram is initially expressed. This is the final expression for the real axis contribution to the Green's function and is the central result of this paper.
We notice that all the arguments of the propagators are now written in an arbitrary representation.
We still have the sum over $l$, which runs over $2 n$ representations, but we only use representation $t^{(l)}$ to determine the sign of the Fermi function $f(b^{(l)}_\beta y_\beta)$ and to figure out whether pseudo-particle Green's function needs to be conjugated or we need to take the real part $G^{(a_\alpha t^{(l)}_{j,\alpha})}_j$. Furthermore, $a_\alpha t^{(l)}_{j,\alpha}$ when not zero, has the same sign for all $j$, and its value is $a_\alpha b_\alpha^{(l)}$, because $t^{(l)}_{j,\alpha}=b_\alpha^{(l)} c^{(l)}_{j,\alpha}$ where $c^{(l)}_{j,\alpha}$ can only be 0 or 1.
As a consequence, the sign of each term in the sum ($a_\alpha b_\alpha^{(l)}$) is easy to determine: If the $l$ term has any pseudo-particle Green's function propagator conjugated, then the sign is negative as all pseudo-particle Green's functions in this term are either conjugated or we need to take their real part. 
On the other hand, if any pseudo-particle Green's function in the $l$ term is non conjugated, then none is conjugated, and the sign is positive.

Finally, we state the Feynman rules for calculating the electron Green’s function on the real axis. 
Comparing the real-axis expression in Eq.~\ref{Eq:reG} with the imaginary-axis expression in Eq.~\ref{imagaG}, we find that the following substitutions are required:
\begin{itemize}
\item the hybridization function is evaluated on the real axis by making the substitution  $\Delta_\beta(a_\beta i\nu_\beta)\rightarrow A^c_\beta(a_\beta y_\beta)=-\frac{1}{\pi}\textrm{Im}(\Delta_\beta(a_\beta y_\beta))$

\item the Matsubara sums over the hybridization frequencies $i\nu_\beta$ are replaced by integrals over the real variables $y_\beta$, while the sum over $i\omega$ is replaced by an integral over $\omega$.
  
\item We need to sum over $2 n$ terms in which one of the pseudo-particle propagators $G_l(i\omega+\cdots)$ is replaced by its spectral function $\widetilde{A}_l(\omega+\lambda+\cdots)$ while the rest $2 n-1$ pseudo-particle Green's functions are substituted with the real part, the retarded, or the advanced Green's function on the real axis. This is determined by the argument $a_\alpha t^{(l)}_{j,\alpha}$, which can be 0, 1, or -1, i.e., $G^{(a_\alpha t^{(l)}_{j,\alpha})}_j(\omega+\lambda+\cdots)$.
The overall sign $a_\alpha b^{(l)}_\alpha$ of the term is positive when retarded $G_j$ appear, and negative when advanced $G_j$ appear.
Note that a mixed term containing both advanced and retarded $G_j$ is not possible. 


\item Finally, each term requires a different set of $n-1$ Fermi functions, which are given by $b^{(l)}_\beta$ in representation $l$, i.e., $\prod_{\beta\ne\alpha}f(b^{(l)}_\beta y_\beta)$. The same set of Fermi functions is required to compute pseudo-particle self-energy $\Sigma_l$ on the real axis.
\end{itemize}

We next apply these rules to our example in Fig.~\ref{SFig2}. We start with the modified generating functional 
$\overline{\Phi}$, defined in Eq.~\ref{modified}.
We then construct $2 n=6$ terms in which exactly one $G$ is replaced by its spectral function $\widetilde{A}$ and some are replaced by their real part. For example, when we compute $\cG_1(-y_1)$, and we concentrate on $(l)=(0)$, we have $a_1=-1$ and $t^{(0)}_{:,1}=(0,0,-1,-1,-1,-1)$, hence $a_1 t^{(0)}_{:,1}=(0,0,1,1,1,1)$. The zeroth component $G_0$ is replaced by the spectral function $\widetilde{A}_0$, while component 1 vanishes, and hence requires $G_1$ to be replaced by the real part $\textrm{Re}G_1$. The rest of the Green's functions should be retarded. The sign of the term is positive. In the second term $(l)=1$ we have the same $a_1 t^{(1)}_{:,1}=(0,0,1,1,1,1)$, hence only for $G_0$ we take the real part, while $G_1$ is replaced by $\widetilde{A}_1$. The next four terms have the same $a_1 t^{(2)}_{:,1}=(-1,-1,0,0,0,0)$, which requires complex conjugated $G_0$ and $G_1$, while the rest of $G$'s are replaced by their real part, except for one, which requires the spectral function. The explicit form is:
\begin{eqnarray}
 \cG_1(-y_1)\braket{Q}=&& 
\left(\frac{\delta \overline{\Phi}}{\delta A^c_1(-y_1)}\right)
  \left(
  f(-y_2)f(y_3)  \frac{\widetilde{A}_0(\omega+y_1+y_2) G'_1(\omega+y_1)}{G_0(\omega+y_1+y_2) G_1(\omega+y_1)}+
  f(y_2)f(y_3) \frac{G'_0(\omega+y_1+y_2)\widetilde{A}_1(\omega+y_1)}{G_0(\omega+y_1+y_2) G_1(\omega+y_1)}
  \right)\nonumber\\
&-&\left(\frac{\delta \overline{\Phi}}{\delta A^c_1(-y_1)}\right)^*
  \left(
  f(y_2)f(y_3) \frac{\widetilde{A}_2(\omega)G'_3(\omega+y_3)G'_4(\omega+y_2+y_3)G'_5(\omega+y_2)}{G^*_2(\omega)G^*_3(\omega+y_3)G^*_4(\omega+y_2+y_3)G^*_5(\omega+y_2)}
  \right)\nonumber\\
 &-&\left(\frac{\delta \overline{\Phi}}{\delta A^c_1(-y_1)}\right)^*
\left(  
  f(y_2)f(-y_3)\frac{ G'_2(\omega) \widetilde{A}_3(\omega+y_3)G'_4(\omega+y_2+y_3)G'_5(\omega+y_2)}{G^*_2(\omega)G^*_3(\omega+y_3)G^*_4(\omega+y_2+y_3)G^*_5(\omega+y_2)}
\right)
 \nonumber\\
 &-&\left(\frac{\delta \overline{\Phi}}{\delta A^c_1(-y_1)}\right)^*
\left(    
    f(-y_2)f(-y_3)\frac{ G'_2(\omega)G'_3(\omega+y_3) \widetilde{A}_4(\omega+y_2+y_3) G'_5(\omega+y_2)}{G^*_2(\omega)G^*_3(\omega+y_3)G^*_4(\omega+y_2+y_3)G^*_5(\omega+y_2)}
\right)
\nonumber\\
&-&\left(\frac{\delta \overline{\Phi}}{\delta A^c_1(-y_1)}\right)^*
\left(    
    f(-y_2)f(y_3)\frac{ G'_2(\omega)G'_3(\omega+y_3)G'_4(\omega+y_2+y_3) \widetilde{A}_5(\omega+y_2)}{G^*_2(\omega)G^*_3(\omega+y_3)G^*_4(\omega+y_2+y_3)G^*_5(\omega+y_2)}
\right)
\end{eqnarray}
The Fermi functions are constructed from quantities $b^{(l)}_\alpha$ listed above.
For the second contribution to the electron Green's function $\alpha=2$, we again need to generate 6 terms, which are:
\begin{eqnarray}
 \cG_2(y_2)\braket{Q}=&& 
\left(\frac{\delta \overline{\Phi}}{\delta A^c_2(y_2)}\right)
  \left(
  f(-y_1)f(y_3)  \frac{\widetilde{A}_1(\omega+y_1) G'_2(\omega)G'_3(\omega+y_2)}{G_1(\omega+y_1) G_2(\omega)G_3(\omega+y_2)}+
  f(y_1)f(y_3) \frac{G'_1(\omega+y_1)\widetilde{A}_2(\omega)G'_3(\omega+y_2)}{G_1(\omega+y_1) G_2(\omega)G_3(\omega+y_2)}
  \right)\nonumber\\
&+&\left(\frac{\delta \overline{\Phi}}{\delta A^c_2(y_2)}\right)
  \left(
  f(y_1)f(-y_3)  \frac{G'_1(\omega+y_1)G'_2(\omega)\widetilde{A}_3(\omega+y_2)}{G_1(\omega+y_1) G_2(\omega)G_3(\omega+y_2)}
  \right)\nonumber\\
&-&\left(\frac{\delta \overline{\Phi}}{\delta A^c_2(y_2)}\right)^*
  \left(
    f(-y_1)f(y_3) \frac{\widetilde{A}_0(\omega+y_1+y_2)G'_4(\omega+y_2+y_3)G'_5(\omega+y_2)}{G^*_0(\omega+y_1+y_2)G^*_4(\omega+y_2+y_3)G^*_5(\omega+y_2)}
\right)\nonumber\\
 &-&\left(\frac{\delta \overline{\Phi}}{\delta A^c_2(y_2)}\right)^*
\left(    
f(y_1)f(-y_3)
\frac{G'_0(\omega+y_1+y_2)\widetilde{A}_4(\omega+y_2+y_3)G'_5(\omega+y_2)}{G^*_0(\omega+y_1+y_2)G^*_4(\omega+y_2+y_3)G^*_5(\omega+y_2)}        
\right)
\nonumber\\
 &-&\left(\frac{\delta \overline{\Phi}}{\delta A^c_2(y_2)}\right)^*
\left(  
     f(y_1)f(y_3)
\frac{G'_0(\omega+y_1+y_2)G'_4(\omega+y_2+y_3)\widetilde{A}_5(\omega+y_2)}{G^*_0(\omega+y_1+y_2)G^*_4(\omega+y_2+y_3)G^*_5(\omega+y_2)}         
\right)
\end{eqnarray}

\end{widetext}

\end{document}